\documentclass[11pt]{article}

\usepackage[a4paper,margin=2.5cm]{geometry}
\usepackage[T1]{fontenc}
\usepackage[utf8]{inputenc}
\usepackage{times}
\usepackage{microtype}
\usepackage{amsmath}
\usepackage{graphicx}
\usepackage{booktabs}
\usepackage{longtable}
\usepackage{float}
\usepackage[font=small,labelfont=bf]{caption}
\usepackage{authblk}
\usepackage{xcolor}
\usepackage[numbers,super,sort&compress]{natbib}
\usepackage[colorlinks=true,allcolors=blue]{hyperref}
\usepackage{url}

\title{\bfseries All four leading LLMs talk more than they listen to\\ personality-verified synthetic help-seekers}

\author[1]{Pablo A. Fonseca}
\author[1,2]{Raquel Rodríguez-Carvajal\thanks{Corresponding author: \texttt{raquel.rodriguez@uam.es}}}
\author[1]{Rafael A. Calvo}

\affil[1]{Dyson School of Design Engineering, Imperial College London, London, United Kingdom}
\affil[2]{Departamento de Psicología Biológica y de la Salud, Universidad Autónoma de Madrid, Madrid, Spain}

\date{}

\begin{document}
\maketitle

\begin{abstract}
\noindent
Large language models are increasingly consulted at moments of distress, yet single-turn benchmarks neither test sustained exchanges nor distinguish between users. We built a personality-aware evaluation in which four widely used models advised several synthetic help-seekers, each given a psychometrically specified profile, in an acute crisis: a caregiver learning of a relative's dementia diagnosis. Auditors blind to the profile prompt recovered the specified bands from dialogue alone with high agreement on every instrument (ICC(2,4) = 0.91; 0.79--0.96 by instrument; band-score $r$ = 0.78), as expected for the Big Five but equally for coping style, coping self-efficacy, resilience and reactance, which the lexical approach never covered. Such evaluation therefore reaches beyond the Five Factor Model to motivational, regulatory and self-appraisal dispositions. The four models were not distinguishable on emotion stabilisation and failed alike, sharing three modes: verbosity, a talk-to-listen ratio above one, and problem-solving before the situation had been explored.
\end{abstract}

\section{Introduction}

People increasingly turn to large language models (LLMs) for health information and support, and they do so most consequentially at moments of distress, when a person faces high uncertainty, low controllability and high emotional intensity, and seeks certainty, information and reassurance. These moments cluster around medical encounters: in the anticipation of a consultation and, especially, in its immediate aftermath, when difficult news has not yet been assimilated and the drive to reduce uncertainty is most acute. Uncertainty of this kind is itself a driver of distress and of information-seeking behaviour\cite{brashers2001} and it is precisely at these moments that formal support is scarcest. Yet the impact of LLMs in these high-use, high-stakes moments remains largely unstudied: most evaluations assess factual accuracy on single-turn queries rather than how these systems behave over a sustained, emotionally charged conversation.

This gap matters because a crisis reshapes the very capacities that a support conversation must accommodate. Acute stress narrows attention and reduces the working-memory and executive resources available to process new information\cite{thomas2024,arnsten2009,shields2016}. A person in this state has a low attentional budget: they can absorb little at once, and evidence-based crisis support therefore prioritises listening and stabilisation before information-giving and problem-solving\cite{roberts2005,wang2024}. Whether LLMs, which tend to be verbose and quick to advise, respond in a way that fits this constraint is an open and pressing question, and one that a single-shot accuracy benchmark cannot answer.

To examine this question, we require a crisis context that is both representative of acute distress and tractable enough that a single precipitating event can be held constant across all conversations, so that any observed variation in the interaction can be attributed to differences in psychological profile rather than to the stressor itself. Family caregivers of people with dementia carry one of the heaviest and least supported mental-health burdens in medicine. Dementia already affects tens of millions of people worldwide, and most of their care is provided informally by relatives\cite{who2021}; among spousal caregivers, roughly 60\% develop a clinically significant depressive or anxiety disorder within two years of taking on the role\cite{joling2015}. Formal support is scarce, stigmatised and hard to reach at the moments of acute distress, such as a new diagnosis, when it is most needed, and caregivers are already turning to large language models to fill the gap, posing real questions about a relative's dementia to systems such as ChatGPT\cite{aguirre2024}. Yet the same systems can fail in high-stakes moments, expressing stigma toward people with mental-health conditions and, in documented cases, responding to expressions of suicidal ideation by supplying means rather than support\cite{moore2025}. Understanding when, for whom, and how LLM-based support helps or harms caregivers in crisis is therefore an urgent scientific priority. Some are addressing it through participatory approaches\cite{shi2026}, but data-driven methods should complement them.

Seeking healthcare advice is among the most common uses of these systems, delivered through conversational interfaces\cite{costagomes2026}. Yet little is known about their quality, and the quality changes at the fast pace of model development. Evidence suggested that they achieved the equivalent of a passing score for a third-year medical student\cite{gilson2023}, but they failed to cultivate rapport or follow medical ethics protocols such as support of autonomy\cite{safranek2023}. Evaluations of the impact of LLMs in improving people's health are fewer, more challenging to achieve, and much more important.

On the one hand this includes the safety evaluations of LLMs. These rely heavily on static benchmarks which are rapidly rendered obsolete when the model is trained on them, and generally fail to capture the dynamic, interactive nature of real clinical encounters\cite{pan2025}. Domain-specific efforts have begun to close part of this gap: MindBench.ai\cite{dwyer2025}, for instance, evaluates against mental-health specific and clinically grounded criteria. Such benchmarks substantially improve on general-purpose leaderboards, but they still characterise the systems through the response to discrete prompts and leave open how an Advisor's conduct might evolve across a sustained, emotionally charged interaction. Furthermore, recent evidence demonstrates that while LLMs may pass static safety filters, their behavioural adherence degrades significantly under persistent, multi-turn dialogue, especially in high-stakes health scenarios where users express distress\cite{hussain2026}.

But besides being safe and not causing harm, LLMs should add value to the user's health. The impact on well-being indicators of some LLMs has been assessed in a number of randomised controlled trials\cite{omar2024}, and the first such trial of an expert-fine-tuned generative-AI chatbot reported significant reductions in depression, anxiety and eating-disorder symptoms relative to a waitlist control\cite{heinz2025}. But in general, reviews of applications of LLMs in mental health\cite{bucher2025,guo2024} have highlighted a lack of evidence on the well-being and support outcomes and safety of LLMs.

One aspect known to affect both safety, that is, preventing negative outcomes, and efficacy, that is, promoting positive outcomes, is the working alliance between the person seeking help and the source of support. In human psychotherapy, this alliance mediates the link between pretreatment personality and symptom reduction during treatment for major depressive disorder\cite{kushner2016}. Whether a comparable relational dynamic emerges when the source of support is an LLM Advisor, rather than a human clinician, remains an open empirical question. In human therapy, the strength of this alliance depends in part on how well the clinician adapts their communication style to the patient's psychological profile; an analogous sensitivity to individual differences may shape how well an LLM Advisor supports a given user.

This therapeutic alliance requires that the therapist differentiates among patients' psychological profiles, which shapes treatment outcomes and the broader treatment process\cite{kushner2016}, and these are increasingly recognised as central to case conceptualisation and treatment planning\cite{bucher2019,bagby2016}. Dispositional factors are associated with outcome, treatment barriers, and therapeutic process --- in a meta-analysis of 99 studies these associations were small but consistent across domains and outcomes\cite{kushner2016} --- and some of these effects are reflected in, but not exhausted by, therapeutic alliance\cite{bucher2019,eells2022,zilchamano2022}. Studies also suggest that LLM behaviour can be adapted to simulate personality traits and mental health states\cite{vu2026}. Taken together, this suggests that evaluation of an LLM Advisor should be sensitive to the Advisee's psychological profile: if profiles shape how support is received, an evaluation instrument must first be able to instantiate and recover them. Whether Advisor quality is in fact moderated by profile is a further question, which a design with a small number of deterministic profiles can raise but not settle. Existing evaluations, however, are ill-suited to this task. They are predominantly single-turn and static, they treat the model in isolation rather than as one participant in an evolving relationship, and because probing real crises with real people is neither safe nor scalable, they rarely test how support unfolds across a sustained, emotionally charged exchange with a psychologically specific help-seeker. What is missing is a way to examine the interaction itself: how a given help-seeker, a given model, and an independent judge of quality behave together, dynamically and at scale, without exposing anyone to harm. This motivates an approach built on three separable roles --- a synthetic Advisee who seeks help, an Advisor who provides it, and an Auditor who evaluates it --- so that realistic, adversarial clinical conditions can be simulated safely and the contribution of each role isolated. Crucially, these three roles are defined by function rather than by any particular model: in principle any LLM, or a human, can occupy each one, and in the studies below the same models appear in different roles across experiments.

In this paper we introduce the AAA framework and demonstrate it on the acute-crisis moment following the disclosure of a dementia diagnosis to a close relative. Seven synthetic Advisees, whose psychological profiles are carefully specified along dimensions selected because they are among the strongest predictors of caregivers' mental-health trajectory (personality, coping, self-efficacy, reactance and resilience), turn to one of four widely available LLMs acting as Advisors (GPT-4.1, Claude-Sonnet-4.5, Mistral-Medium and DeepSeek-V3.2). Every conversation is then rated by an independent LLM Auditor, blind to the psychometric specification, across three complementary outcome domains: emotional stabilisation, satisfaction of basic psychological needs, and quality of the support behaviour. For this last domain we introduce a new behavioural counter, grounded in the clinical literature on crisis intervention, motivational interviewing and evidence-based therapeutic relationships, that flags reactance-producing expressions, process failures and boundary and scope violations.

Our contribution is thus a reusable, safe and personality-aware framework for evaluating LLM-based support in acute crisis: a role-separated architecture with synthetic Advisees whose psychological profile can be prescribed and verified, and an Auditor panel augmented by a new clinically grounded behavioural counter.

Whether a specified profile can be recovered from dialogue is not equally well established across the dimensions we prescribe, and this is not a purely technical point. That a person's standing on the Big Five is expressed in, and recoverable from, natural language is well supported: these dimensions rest on the lexical hypothesis, and language-based assessment of them converges with self-reports and informant reports\cite{goldberg1990,park2015}. It also holds when the speaker is a model rather than a person, though so far only for that inventory: Big Five measures administered to LLMs yield reliable and valid scores in sufficiently scaled, instruction-tuned models, and the levels expressed in their output can be shaped along the intended dimensions\cite{serapiogarcia2025}. Both literatures stop at the Five Factor Model, which summarises descriptive style at the level of traits and was never designed to represent motivational dispositions, regulatory and coping strategies, or appraisals of one's own capacity: a psychology of the stranger rather than of the real person\cite{mcadams1992}. For dispositional constructs of this kind much less is known. Coping style, coping self-efficacy, resilience and reactance are general characteristics of a person rather than features of a situation, but they are appraisals of one's own capacity and motivational state rather than descriptions of observable style; they are also among the stronger predictors of caregiver trajectory, which is why they are the ones we prescribe. They are assessed almost exclusively by questionnaire, and there is little evidence on whether they surface in conversation at all, still less in a form an independent judge could score. The asymmetry has particular force when both the Advisee and the Auditor are language models. What such a model represents about persons is built from text, so dispositional content that is lexically encoded is precisely the content it is best placed both to express and to detect, while appraisals that speakers rarely put into words have no comparable lexical footprint to draw on. The expectation going in was therefore a marked gradient: the lexically derived dimensions recoverable, and the appraisal-based instruments recoverable far less well, or not at all. How far that expectation holds sets a ceiling on how much of a psychological profile any personality-aware evaluation can act upon.

Our four studies aim to answer the following questions:

\begin{itemize}
\item Can a synthetic Advisee be given a specified psychological profile that it portrays consistently in dialogue?
\item Can such a profile be recovered by an independent Auditor from the dialogue alone, without access to the specification, thereby establishing the evaluation as personality-aware?
\item Does the quality of support differ across the four Advisors under this personality-aware evaluation, and can the behavioural counter identify the specific conducts that account for those differences?
\end{itemize}

\section{Results}

\subsection{Overview}

The four studies address three research questions. The first two, that is, whether a specified psychological profile can be reproduced in a synthetic Advisee, and whether it can be recovered by an independent Auditor from the resulting conversation, are examined jointly in Study 1, and are a precondition for interpreting the remaining results: unless the Advisee reliably carries the specified profile into the interaction, differences among Advisors in Studies 2 to 4 cannot be safely attributed to profile variation rather than to noise on the Advisee side. We distinguish the profile (the psychometric specification of an Advisee, given in Supplementary Table 2) from the instantiating model (the LLM that enacts it in dialogue). Five distinct profiles are specified, two of which are additionally instantiated as matched pairs, one varying only gender and one varying only age, yielding seven Advisees in total. In Study 1 each Advisee is instantiated with four different models (GPT-4.1, Claude-Sonnet-4.5, Mistral-Medium, DeepSeek-V3.2) in order to identify which reproduces the specified profile most faithfully; in Studies 2 to 4 the instantiating model is fixed to GPT-4.1, so that the profile varies while the Advisee-side model is held constant. The third question asks whether the quality of the Advisor differs across models and along which dimensions; this is examined in Studies 2 to 4, each centred on a distinct caregiver outcome in the acute situation: emotional stabilisation, measured through affect and perceived stress (Study 2); the support of basic psychological needs, framed within self-determination theory (Study 3); and the quality of the support provided and the Advisor's observable conduct during the interaction (Study 4).

\subsection{Study 1 --- Personality traits in synthetic Advisees}

\noindent\textbf{Advisee:} the seven specified synthetic Advisees, each instantiated in turn with GPT-4.1, Claude-Sonnet-4.5, Mistral-Medium and DeepSeek-V3.2; the instantiating model is the manipulated factor. \textbf{Advisor:} fixed to GPT-4.1 with a minimal system prompt, held constant so that variation is attributable to the Advisee side. \textbf{Auditor:} an ensemble of all four models, blind to the psychometric specification, scoring the full conversation; reported values are means across the ensemble. This yields 28 conversations in the specified-profile arm (7 profiles $\times$ 4 instantiating models) and 28 in the no-profile baseline arm, 56 in total.

This study addresses the first two research questions jointly: whether a specified profile can be reproduced in a synthetic Advisee, and whether it can be recovered by an independent Auditor from the conversation alone. Each reported score is the mean of four independent Auditors, effectively using them in an ensemble configuration. We used an ensemble here because recovering a specified psychological profile is the measurement on which all subsequent studies depend, so averaging across heterogeneous Auditors guards against the idiosyncrasy or bias of any single model; in Studies 2 to 4, where the Advisee is fixed, a single higher-capability Auditor (gpt-5.4) was used instead.

For each synthetic Advisee conversation, the trait Auditors inferred a score from 1 to 5 for the ten constructs (three NEO-FFI traits, four CHIP coping styles, ResQ-Care resilience, CSES-8 self-efficacy and Hong reactance) for the seven synthetic Advisee instances. We broke the bands into three levels --- low [1--2.33] (centre 1.67), medium (2.33--3.66] (centre 3.00) and high [3.66--5] (centre 4.34) --- for evaluating the resulting scores. In Table~\ref{tab:fidelity} we report $N$ as the number of construct instances (10 constructs $\times$ 7 synthetic Advisees), the band-score Pearson correlation, the Spearman rank correlation, in-band accuracy (the percentage of times the score falls in the target band) and the mean absolute error from the target band centre.

Recoverability was not expected to be uniform across the ten constructs. In the person-perception literature, agreement between independent judges is consistently higher for dispositions whose expression is directly observable and lower for those that are internal or evaluative, a regularity usually described as trait visibility: observability raises interjudge agreement while evaluativeness lowers it\cite{john1993,ashton2025}. Observability there refers to conduct available to an onlooker in the broad sense, including facial expression, gesture, activity level and the company a person keeps. The channel available to our Auditors is narrower: they work from a transcript, so the only observable conduct is verbal behaviour, meaning what the Advisee says and how it is said. We therefore expected the NEO-FFI dimensions, whose content is expressed continuously in the register, lexicon and content of speech, to be recovered more reliably than coping self-efficacy and reactance, which are appraisals of one's own capacity and motivational state and need not surface in the wording of a twenty-turn conversation at all. We report agreement by instrument for this reason, rather than only in aggregate.

\begin{table}[H]
\centering
\caption{Recovery of specified Advisee profiles by the Auditor ensemble, by instantiating model and condition. $N$ is the number of construct instances (10 constructs $\times$ 7 Advisees). Scores are means across the four-model Auditor ensemble. Correlations are between the recovered score and the centre of the specified band. In-band accuracy is the proportion of instances whose recovered score falls within the specified band. Lower MAE indicates closer agreement with the specified band centre. 95\% confidence intervals are obtained by a cluster bootstrap ($B = 10{,}000$) that resamples Advisees with replacement.}
\label{tab:fidelity}
\resizebox{\textwidth}{!}{\footnotesize\setlength{\tabcolsep}{4pt}
\begin{tabular}{llrrlrrlr}
\toprule
\textbf{Provider} & \textbf{Condition} & \textbf{N} & \textbf{r} & \textbf{r 95\% CI} & \textbf{Spearman} & \textbf{In-band acc.} & \textbf{acc 95\% CI} & \textbf{MAE} \\
\midrule
GPT-4.1 & Specified profile & 70 & 0.779 & [0.675, 0.851] & 0.771 & 0.629 & [0.543, 0.714] & 0.551 \\
GPT-4.1 & No-profile baseline & 70 & 0.41 & [0.247, 0.580] & 0.412 & 0.529 & [0.443, 0.614] & 0.827 \\
Claude-Sonnet-4.5 & Specified profile & 70 & 0.744 & [0.640, 0.826] & 0.742 & 0.586 & [0.514, 0.657] & 0.601 \\
Claude-Sonnet-4.5 & No-profile baseline & 70 & 0.374 & [0.254, 0.475] & 0.343 & 0.471 & [0.386, 0.543] & 0.862 \\
Mistral-Medium & Specified profile & 70 & 0.689 & [0.544, 0.794] & 0.694 & 0.557 & [0.429, 0.657] & 0.651 \\
Mistral-Medium & No-profile baseline & 70 & 0.409 & [0.280, 0.498] & 0.389 & 0.486 & [0.371, 0.600] & 0.818 \\
DeepSeek-V3.2 & Specified profile & 70 & 0.715 & [0.588, 0.806] & 0.697 & 0.5 & [0.414, 0.571] & 0.644 \\
DeepSeek-V3.2 & No-profile baseline & 70 & 0.289 & [0.140, 0.431] & 0.281 & 0.4 & [0.343, 0.457] & 0.926 \\
\bottomrule
\end{tabular}}
\end{table}

To confirm that recovery reflects the enacted profile rather than the shared scenario, biographical detail or Auditor priors, we repeated the procedure with the psychometric specification removed from the Advisee prompt and scored the resulting conversations against the same specifications as a placebo control. Demographics, scenario and task were byte-identical between the two arms, and both were scored by the same specification-blind Auditors. Recovery fell for every instantiating model, and the profile effect was reliable throughout (Table~\ref{tab:persona}). Baseline recovery was not zero ($r$ = 0.29--0.41), indicating that part of the recoverable signal comes from the scenario and demographic context rather than from the specified profile; what the control establishes is that this context cannot account for the additional recovery obtained when a profile is supplied.

\begin{table}[H]
\centering
\caption{Persona effect: the difference in band-score correlation between the specified-profile and no-profile baseline conditions. Intervals and $p$ values come from the paired cluster bootstrap over Advisees.}
\label{tab:persona}
\footnotesize\setlength{\tabcolsep}{4pt}
\begin{tabular}{lrlrl}
\toprule
\textbf{Provider} & \textbf{$\Delta$r (profile $-$ baseline)} & \textbf{95\% CI} & \textbf{p (boot)} & \textbf{CI excludes 0} \\
\midrule
GPT-4.1 & 0.369 & [0.225, 0.489] & 0.0 & yes \\
Claude-Sonnet-4.5 & 0.37 & [0.294, 0.431] & 0.0 & yes \\
Mistral-Medium & 0.28 & [0.221, 0.345] & 0.0 & yes \\
DeepSeek-V3.2 & 0.425 & [0.276, 0.570] & 0.0 & yes \\
\bottomrule
\end{tabular}
\end{table}

Agreement among the four Auditors was assessed with two-way random-effects intraclass correlations using absolute agreement (Table~\ref{tab:icc}). Reliability of the reported ensemble mean was ICC(2,4) = .906, while a single Auditor would achieve only ICC(2,1) = .706, indicating that the ensemble contributes materially to the stability of the estimate. As anticipated, agreement was highest for the NEO-FFI dimensions (ICC(2,1) = .871; ICC(2,4) = .964) and lowest for reactance (.486; .791) and CSES-8 self-efficacy (.527; .817), with CHIP (.545; .827) and ResQ-Care (.626; .870) in between, so a single Auditor is not sufficient for the less observable constructs.

\begin{table}[H]
\centering
\caption{Agreement among the four Auditors, as two-way random-effects intraclass correlations with absolute agreement. ICC(2,1) is the reliability of a single judge; ICC(2,4) is the reliability of the ensemble mean, which is what the study reports.}
\label{tab:icc}
\footnotesize\setlength{\tabcolsep}{4pt}
\begin{tabular}{lrrr}
\toprule
\textbf{Subset} & \textbf{N targets} & \textbf{ICC(2,1) single judge} & \textbf{ICC(2,4) panel mean} \\
\midrule
All constructs & 280 & 0.706 & 0.906 \\
NEO-FFI & 84 & 0.871 & 0.964 \\
CHIP & 112 & 0.545 & 0.827 \\
ResQ-Care & 28 & 0.626 & 0.87 \\
CSES-8 & 28 & 0.527 & 0.817 \\
Reactance & 28 & 0.486 & 0.791 \\
\bottomrule
\end{tabular}
\end{table}

Because GPT-4.1 serves both as an instantiating model and as one of the four Auditors, we recomputed the recovery analysis with each Auditor removed in turn (Table~\ref{tab:loo}): GPT-4.1 remained first in all five panel configurations, including when it was itself excluded ($r$ = .768 against .779 with the full panel). The lower positions are not stable: Mistral-Medium and DeepSeek-V3.2 exchange third and fourth place depending on which Auditor is removed, so only the identity of the best instantiating model should be read as robust to Auditor composition.

\begin{table}[H]
\centering
\caption{Fidelity recomputed with each Auditor removed in turn, testing whether the ranking depends on any one Auditor.}
\label{tab:loo}
\footnotesize\setlength{\tabcolsep}{4pt}
\begin{tabular}{lllll}
\toprule
\textbf{Judge panel} & \textbf{GPT-4.1} & \textbf{Claude-Sonnet-4.5} & \textbf{Mistral-Medium} & \textbf{DeepSeek-V3.2} \\
\midrule
All four (as reported) & 0.779 (\#1) & 0.744 (\#2) & 0.689 (\#4) & 0.715 (\#3) \\
excluding gpt-4.1 & 0.768 (\#1) & 0.723 (\#2) & 0.681 (\#3) & 0.681 (\#4) \\
excluding claude-sonnet-4-5 & 0.750 (\#1) & 0.706 (\#2) & 0.686 (\#3) & 0.672 (\#4) \\
excluding mistral-medium-2505 & 0.782 (\#1) & 0.753 (\#2) & 0.677 (\#4) & 0.730 (\#3) \\
excluding DeepSeek-V3.2 & 0.794 (\#1) & 0.762 (\#2) & 0.700 (\#4) & 0.729 (\#3) \\
\bottomrule
\end{tabular}
\end{table}

Self-preference was not uniform across Auditors (Table~\ref{tab:selfpref}): GPT-4.1 and Mistral-Medium scored their own model's conversations slightly higher than others', Claude-Sonnet-4.5 was neutral, and DeepSeek-V3.2 scored its own model lower than others'. Judges scoring conversations generated by their own model reached a mean $r$ of .694, against .652 for other models. Each $r$ is computed over 70 construct instances (10 constructs $\times$ 7 Advisees) per Auditor $\times$ instantiating-model cell; comparing the four diagonal cells against the twelve off-diagonal ones gives .694 versus .652 ($t(14) = 0.69$, $p = .50$; permutation $p = .21$). The difference is small and not statistically distinguishable, but it runs in the direction of self-preference and the design has little power to detect an effect of this size, so it should be read as an absence of evidence for a family effect rather than as evidence against one.

\begin{table}[H]
\centering
\caption{Cross-tabulation of Auditor (rows) against instantiating model (columns), as band-score correlations. Diagonal cells are Auditors scoring their own model.}
\label{tab:selfpref}
\resizebox{\textwidth}{!}{\footnotesize\setlength{\tabcolsep}{4pt}
\begin{tabular}{lrrrr}
\toprule
\textbf{Judge} & \textbf{GPT-4.1} & \textbf{Claude-Sonnet-4.5} & \textbf{Mistral-Medium} & \textbf{DeepSeek-V3.2} \\
\midrule
GPT-4.1 & 0.765 & 0.748 & 0.68 & 0.729 \\
Claude-Sonnet-4.5 & 0.782 & 0.709 & 0.655 & 0.683 \\
Mistral-Medium & 0.657 & 0.544 & 0.677 & 0.349 \\
DeepSeek-V3.2 & 0.699 & 0.662 & 0.641 & 0.623 \\
\bottomrule
\end{tabular}}
\end{table}

In paired cluster-bootstrap comparisons (Table~\ref{tab:pairwise}), GPT-4.1, Claude-Sonnet-4.5 and DeepSeek-V3.2 were not reliably separable (GPT-4.1 $-$ DeepSeek-V3.2 $\Delta r$ = 0.064 [$-$0.039, 0.207], $p$ = .31; GPT-4.1 $-$ Claude-Sonnet-4.5 $\Delta r$ = 0.035 [0.000, 0.071], $p$ = .047), while all three exceeded Mistral-Medium. GPT-4.1 held first place under every leave-one-Auditor-out panel configuration, and was selected as the instantiating model for Studies 2 to 4 on that basis; any of the three would have served.

\begin{table}[H]
\centering
\caption{Paired provider differences in the specified-profile condition. $\Delta r$ is the difference in band-score correlation between providers, with 95\% intervals from the paired cluster bootstrap using the same resampled Advisee draws for both providers; $p$ is the two-sided bootstrap proportion.}
\label{tab:pairwise}
\footnotesize\setlength{\tabcolsep}{4pt}
\begin{tabular}{lrlrl}
\toprule
\textbf{Comparison} & \textbf{$\Delta$r} & \textbf{95\% CI} & \textbf{p (boot)} & \textbf{CI excludes 0} \\
\midrule
GPT-4.1 $-$ Claude-Sonnet-4.5 & 0.035 & [0.000, 0.071] & 0.047 & yes \\
GPT-4.1 $-$ Mistral-Medium & 0.09 & [0.041, 0.154] & 0.0 & yes \\
GPT-4.1 $-$ DeepSeek-V3.2 & 0.064 & [-0.039, 0.207] & 0.309 & no \\
Claude-Sonnet-4.5 $-$ Mistral-Medium & 0.055 & [0.022, 0.109] & 0.0 & yes \\
Claude-Sonnet-4.5 $-$ DeepSeek-V3.2 & 0.029 & [-0.065, 0.148] & 0.616 & no \\
Mistral-Medium $-$ DeepSeek-V3.2 & -0.026 & [-0.149, 0.103] & 0.656 & no \\
\bottomrule
\end{tabular}
\end{table}

\begin{table}[H]
\centering
\caption{Band-score correlation by instrument and instantiating model.}
\label{tab:instrument}
\resizebox{\textwidth}{!}{\footnotesize\setlength{\tabcolsep}{4pt}
\begin{tabular}{lrrrr}
\toprule
\textbf{Instrument} & \textbf{GPT-4.1} & \textbf{Claude-Sonnet-4.5} & \textbf{Mistral-Medium} & \textbf{DeepSeek-V3.2} \\
\midrule
NEO-FFI & 0.822 & 0.84 & 0.802 & 0.686 \\
CHIP & 0.657 & 0.533 & 0.539 & 0.687 \\
ResQ-Care & 0.591 & 0.642 & 0.733 & 0.662 \\
CSES-8 & 0.935 & 0.859 & 0.966 & 0.759 \\
Reactance & 0.784 & 0.862 & 0.386 & 0.687 \\
\bottomrule
\end{tabular}}
\end{table}

\subsection{Study 2 --- Emotion stabilisation}

\noindent\textbf{Advisee:} the seven profiles, instantiating model fixed to GPT-4.1 on the basis of Study 1. \textbf{Advisor:} varied across GPT-4.1, Claude-Sonnet-4.5, Mistral-Medium and DeepSeek-V3.2; the manipulated factor for Studies 2 to 4. \textbf{Auditor:} a single higher-capability model (gpt-5.4), administering the PSS and PANAS with access only to the Advisee's first two and last two turns, so that pre--post change is inferred from the Advisee's own language rather than from the Advisor's. This yields 28 conversations (7 profiles $\times$ 4 Advisors), the same corpus analysed in Studies 3 and 4 under different instruments.

Two directional hypotheses were specified in advance: that the encounter reduces negative affect, and that it reduces perceived stress, both of which are targets of crisis stabilisation. Positive affect is reported for completeness, but no change is predicted, since raising positive affect is not a goal of stabilisation in the immediate aftermath of a diagnosis. The three outcomes are distinct constructs --- perceived stress is an appraisal of demands relative to resources, whereas positive and negative affect are the two largely independent dimensions of the PANAS --- and each is reported separately whatever the result rather than combined into a single claim. Effect sizes with confidence intervals are the primary quantity and $p$ values are secondary, and differences between Advisors are not inferred from the number of significant within-Advisor tests.

The emotional response in an acute crisis is a transient state that fluctuates with the person's appraisal of the situation rather than a stable trait\cite{kuppens2017,lazarus1984}; the emotion measured here is therefore best understood as an inner state that unfolds over the course of the interaction. We note, however, that our Auditors infer this state from the conversation text rather than measuring the emotional state of a human, and that emotion inference from text carries well-documented limitations\cite{tak2025}; the values reported for synthetic Advisees should thus be read as model-based inferences and not equated with measured human states.

\begin{figure}[H]
\centering
\includegraphics[width=\textwidth]{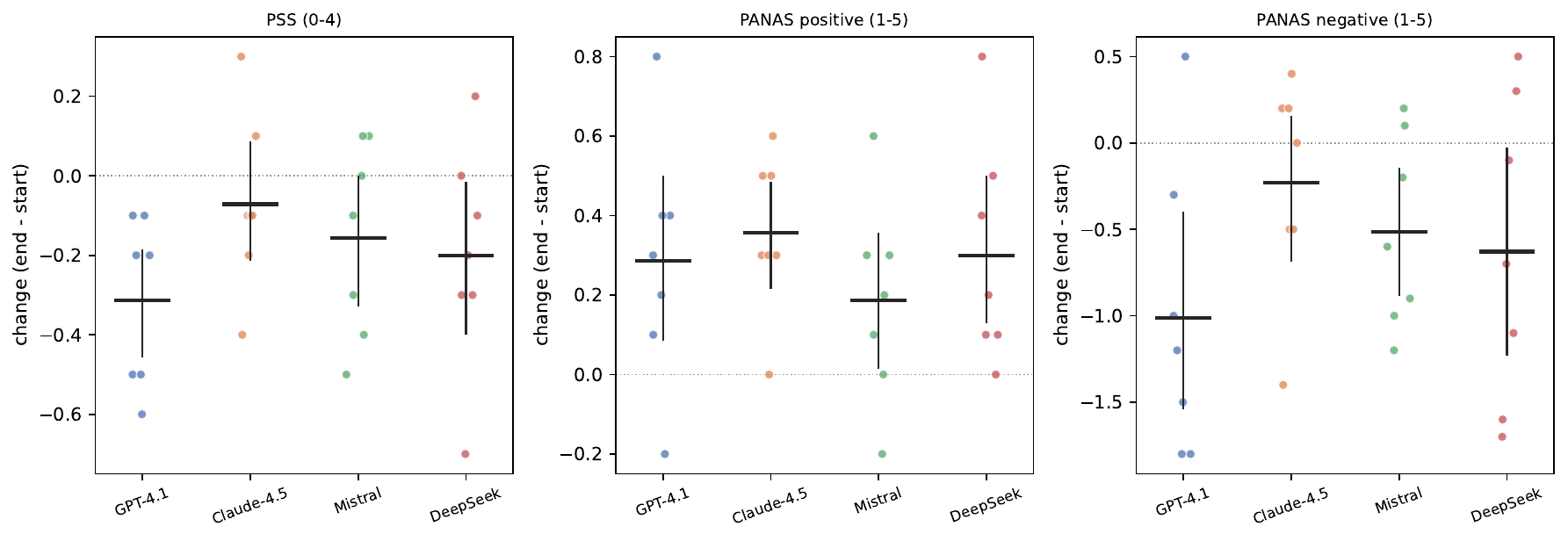}
\caption{Emotional change by Advisor. Each point is one of the seven Advisees; horizontal bars are the mean and the vertical bars are 95\% confidence intervals. For the PSS and PANAS negative, a negative change indicates improvement; for PANAS positive, a positive change indicates an increase.}
\label{fig:emotion}
\end{figure}

\begin{table}[H]
\centering
\caption{Emotional stabilisation: start and end values, change, paired $t$ statistics and corrected $p$ values for the twelve pre--post contrasts. All tests are paired and two-tailed, $t(6)$. Shapiro--Wilk did not reject normality for any contrast (all $p \geq .10$); the Wilcoxon signed-rank $p$ is reported alongside because that test has low power at $n = 7$. Holm correction is applied within instrument, four Advisors per family.}
\label{tab:emotion}
\resizebox{\textwidth}{!}{\footnotesize\setlength{\tabcolsep}{4pt}
\begin{tabular}{llrrrrrrrl}
\toprule
\textbf{Advisor} & \textbf{Measure} & \textbf{Start} & \textbf{End} & \textbf{Delta} & \textbf{t(6)} & \textbf{dz} & \textbf{Wilcoxon p} & \textbf{Holm p} & \textbf{Survives Holm} \\
\midrule
GPT-4.1 & PSS (stress) & 2.29 & 1.97 & -0.31 & -3.93 & -1.49 & 0.0156 & 0.0308 & yes \\
GPT-4.1 & PANAS positive & 1.27 & 1.56 & 0.29 & 2.46 & 0.93 & 0.0625 & 0.0988 & no \\
GPT-4.1 & PANAS negative & 3.2 & 2.19 & -1.01 & -3.17 & -1.2 & 0.0469 & 0.0776 & no \\
Claude-Sonnet-4.5 & PSS (stress) & 2.23 & 2.16 & -0.07 & -0.85 & -0.32 & 0.2969 & 0.4262 & no \\
Claude-Sonnet-4.5 & PANAS positive & 1.21 & 1.57 & 0.36 & 4.75 & 1.8 & 0.0312 & 0.0124 & yes \\
Claude-Sonnet-4.5 & PANAS negative & 3.03 & 2.8 & -0.23 & -0.97 & -0.37 & 0.4688 & 0.3697 & no \\
Mistral-Medium & PSS (stress) & 2.24 & 2.09 & -0.16 & -1.7 & -0.64 & 0.25 & 0.3318 & no \\
Mistral-Medium & PANAS positive & 1.23 & 1.41 & 0.19 & 1.93 & 0.73 & 0.125 & 0.1017 & no \\
Mistral-Medium & PANAS negative & 3.01 & 2.5 & -0.51 & -2.45 & -0.93 & 0.1094 & 0.1491 & no \\
DeepSeek-V3.2 & PSS (stress) & 2.21 & 2.01 & -0.2 & -1.87 & -0.71 & 0.1562 & 0.3318 & no \\
DeepSeek-V3.2 & PANAS positive & 1.33 & 1.63 & 0.3 & 2.81 & 1.06 & 0.0312 & 0.0927 & no \\
DeepSeek-V3.2 & PANAS negative & 3.06 & 2.43 & -0.63 & -1.87 & -0.71 & 0.1562 & 0.2206 & no \\
\bottomrule
\end{tabular}}
\end{table}

As predicted, perceived stress fell for every Advisor, though the reduction survives correction for multiplicity only for GPT-4.1. For PANAS-positive, point estimates increase for all four Advisors; only Claude-Sonnet-4.5 survives Holm correction. We did not predict change on this dimension and report it descriptively: an increase in positive affect is not by itself evidence of stabilisation, since raising positive affect is not a target of crisis support. For the reduction of negative emotion, GPT-4.1 and Mistral-Medium show the largest reductions, neither of which survives Holm correction.

\begin{table}[H]
\centering
\caption{Effect sizes with confidence intervals for the twelve pre--post contrasts. $d_z$ is the paired standardised mean difference; intervals are exact, obtained by inverting the noncentral $t$ distribution ($n = 7$).}
\label{tab:effsize}
\footnotesize\setlength{\tabcolsep}{4pt}
\begin{tabular}{lrlrl}
\toprule
\textbf{Instrument · Advisor} & \textbf{dz} & \textbf{95\% CI of dz} & \textbf{p} & \textbf{CI excludes 0} \\
\midrule
PSS · GPT-4.1 & -1.49 & [-2.57, -0.36] & 0.0077 & yes \\
PSS · Claude-Sonnet-4.5 & -0.32 & [-1.07, 0.45] & 0.4262 & no \\
PSS · Mistral-Medium & -0.64 & [-1.45, 0.20] & 0.1392 & no \\
PSS · DeepSeek-V3.2 & -0.71 & [-1.52, 0.15] & 0.1106 & no \\
PANAS+ · GPT-4.1 & 0.93 & [0.00, 1.80] & 0.0494 & yes \\
PANAS+ · Claude-Sonnet-4.5 & 1.8 & [0.54, 3.01] & 0.0031 & yes \\
PANAS+ · Mistral-Medium & 0.73 & [-0.14, 1.55] & 0.1017 & no \\
PANAS+ · DeepSeek-V3.2 & 1.06 & [0.09, 1.98] & 0.0309 & yes \\
PANAS$-$ · GPT-4.1 & -1.2 & [-2.16, -0.18] & 0.0194 & yes \\
PANAS$-$ · Claude-Sonnet-4.5 & -0.37 & [-1.12, 0.42] & 0.3697 & no \\
PANAS$-$ · Mistral-Medium & -0.93 & [-1.80, -0.00] & 0.0497 & yes \\
PANAS$-$ · DeepSeek-V3.2 & -0.71 & [-1.52, 0.15] & 0.1103 & no \\
\bottomrule
\end{tabular}
\end{table}

\begin{table}[H]
\centering
\caption{Normality and multiplicity. Shapiro--Wilk was applied to the seven paired differences behind every contrast; normality was not rejected in any of the twelve (all $p \geq .10$), though the test has low power at $n = 7$, so each contrast is also reported with the distribution-free Wilcoxon signed-rank test. Familywise correction uses Holm within each instrument (four Advisors per family).}
\label{tab:normmult}
\resizebox{\textwidth}{!}{\footnotesize\setlength{\tabcolsep}{4pt}
\begin{tabular}{lrrrrrrrrr}
\toprule
\textbf{Instrument} & \textbf{Advisor} & \textbf{n} & \textbf{Shapiro-Wilk W} & \textbf{Shapiro p} & \textbf{Normality rejected} & \textbf{paired t p} & \textbf{Wilcoxon p} & \textbf{Holm p (within instrument)} & \textbf{Survives Holm} \\
\midrule
PSS & GPT-4.1 & 7 & 0.84 & 0.1 & no & 0.0077 & 0.0156 & 0.0308 & yes \\
PSS & Claude-Sonnet-4.5 & 7 & 0.944 & 0.675 & no & 0.4262 & 0.2969 & 0.4262 & no \\
PSS & Mistral-Medium & 7 & 0.902 & 0.346 & no & 0.1392 & 0.25 & 0.3318 & no \\
PSS & DeepSeek-V3.2 & 7 & 0.963 & 0.845 & no & 0.1106 & 0.1562 & 0.3318 & no \\
PANAS+ & GPT-4.1 & 7 & 0.969 & 0.891 & no & 0.0494 & 0.0625 & 0.0988 & no \\
PANAS+ & Claude-Sonnet-4.5 & 7 & 0.903 & 0.348 & no & 0.0031 & 0.0312 & 0.0124 & yes \\
PANAS+ & Mistral-Medium & 7 & 0.982 & 0.967 & no & 0.1017 & 0.125 & 0.1017 & no \\
PANAS+ & DeepSeek-V3.2 & 7 & 0.913 & 0.418 & no & 0.0309 & 0.0312 & 0.0927 & no \\
PANAS$-$ & GPT-4.1 & 7 & 0.89 & 0.275 & no & 0.0194 & 0.0469 & 0.0776 & no \\
PANAS$-$ & Claude-Sonnet-4.5 & 7 & 0.879 & 0.222 & no & 0.3697 & 0.4688 & 0.3697 & no \\
PANAS$-$ & Mistral-Medium & 7 & 0.921 & 0.473 & no & 0.0497 & 0.1094 & 0.1491 & no \\
PANAS$-$ & DeepSeek-V3.2 & 7 & 0.921 & 0.474 & no & 0.1103 & 0.1562 & 0.2206 & no \\
\bottomrule
\end{tabular}}
\end{table}

Because the same seven Advisees appear in all four arms, the Advisors can be compared directly while blocking on Advisee. Starting levels accordingly did not differ between Advisors on any outcome: perceived stress $F(3,18) = 0.27$, $p = .844$; positive affect $F(3,18) = 1.08$, $p = .382$; negative affect $F(3,18) = 0.59$, $p = .628$, with ranges across Advisors of 0.07, 0.11 and 0.19 scale points. In a one-way repeated-measures analysis with Advisee as the random factor, no omnibus difference between Advisors was detected on any outcome (Table~\ref{tab:omnibus}). Friedman's rank test agreed in each case. Notably, the Advisee accounted for 45--56\% of the variance in emotional change across the three outcomes, more than the Advisor did.

\begin{table}[H]
\centering
\caption{Omnibus comparison of the four Advisors on the three emotion outcomes, blocking on Advisee.}
\label{tab:omnibus}
\resizebox{\textwidth}{!}{\footnotesize\setlength{\tabcolsep}{4pt}
\begin{tabular}{lrrrrrrrrr}
\toprule
\textbf{Outcome} & \textbf{n Advisees} & \textbf{F} & \textbf{df} & \textbf{p} & \textbf{partial eta\^{}2} & \textbf{Friedman chi2} & \textbf{Friedman p} & \textbf{Advisee share of variance} & \textbf{Advisors differ} \\
\midrule
PSS & 7 & 2.36 & 3,18 & 0.1059 & 0.282 & 6.85 & 0.0768 & 53\% & no \\
PANAS+ & 7 & 0.95 & 3,18 & 0.4382 & 0.136 & 2.43 & 0.4887 & 56\% & no \\
PANAS$-$ & 7 & 2.11 & 3,18 & 0.1351 & 0.26 & 5.23 & 0.1558 & 45\% & no \\
\bottomrule
\end{tabular}}
\end{table}

A caution applies regardless of statistical power: in a synthetic setting the Advisee's emotional state is not truly independent of what the Advisor says, since both are generated by the same class of model, and any apparent improvement may partly reflect linguistic mirroring rather than genuine emotional relief. This would only become meaningful if corroborated with human participants.

\subsection{Study 3 --- Basic psychological needs}

\noindent\textbf{Advisee} and \textbf{Advisor} as in Study 2. \textbf{Auditor:} gpt-5.4, scoring the METUX TENS-Life scale with access to the full conversation history including both speakers' turns, since need support and frustration are properties of the exchange rather than of either party alone.

In this study we are concerned with the Advisors' behaviours supporting or frustrating basic psychological needs as identified by self-determination theory (autonomy, competence, relatedness) and measured by the METUX TENS-Life scale\cite{burnell2023,peters2018,chen2015}.

\begin{figure}[H]
\centering
\includegraphics[width=\textwidth]{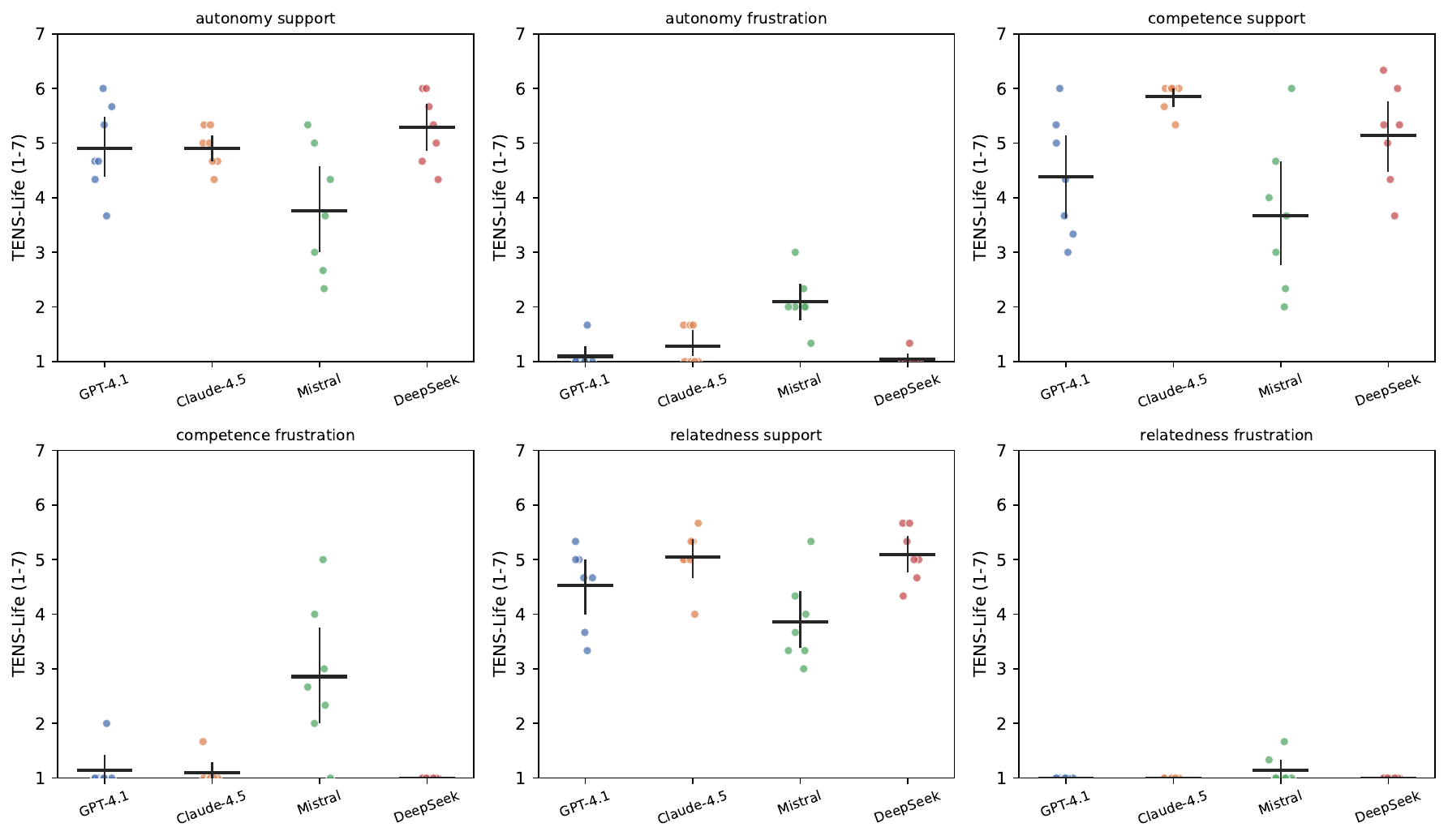}
\caption{TENS-Life scores (1--7) by Advisor. Each point is one of the seven Advisees. TENS separates the support and frustration aspects of the effect on basic psychological needs; larger support and smaller frustration values indicate better Advisor performance. Error bars are 95\% percentile-bootstrap confidence intervals.}
\label{fig:tens}
\end{figure}

\begin{table}[H]
\centering
\caption{Basic psychological needs: autonomy. Values are on a 1--7 scale. Where all seven Advisees scored the scale floor, the bootstrap interval is degenerate ([1.00, 1.00]) and reflects a floor effect rather than high precision.}
\label{tab:autonomy}
\footnotesize\setlength{\tabcolsep}{4pt}
\begin{tabular}{lrlrl}
\toprule
\textbf{Advisor} & \textbf{Support} & \textbf{Support 95\% CI} & \textbf{Frustration} & \textbf{Frustration 95\% CI} \\
\midrule
GPT-4.1 & 4.9 & [4.38, 5.48] & 1.1 & [1.00, 1.29] \\
Claude-Sonnet-4.5 & 4.9 & [4.67, 5.14] & 1.29 & [1.10, 1.57] \\
Mistral-Medium & 3.76 & [3.00, 4.57] & 2.1 & [1.76, 2.43] \\
DeepSeek-V3.2 & 5.29 & [4.86, 5.71] & 1.05 & [1.00, 1.14] \\
\bottomrule
\end{tabular}
\end{table}

\begin{table}[H]
\centering
\caption{Basic psychological needs: competence. Values are on a 1--7 scale. Where all seven Advisees scored the scale floor, the bootstrap interval is degenerate ([1.00, 1.00]) and reflects a floor effect rather than high precision.}
\label{tab:competence}
\footnotesize\setlength{\tabcolsep}{4pt}
\begin{tabular}{lrlrl}
\toprule
\textbf{Advisor} & \textbf{Support} & \textbf{Support 95\% CI} & \textbf{Frustration} & \textbf{Frustration 95\% CI} \\
\midrule
GPT-4.1 & 4.38 & [3.62, 5.14] & 1.14 & [1.00, 1.43] \\
Claude-Sonnet-4.5 & 5.86 & [5.67, 6.00] & 1.1 & [1.00, 1.29] \\
Mistral-Medium & 3.67 & [2.76, 4.67] & 2.86 & [2.00, 3.76] \\
DeepSeek-V3.2 & 5.14 & [4.48, 5.76] & 1.0 & [1.00, 1.00] \\
\bottomrule
\end{tabular}
\end{table}

\begin{table}[H]
\centering
\caption{Basic psychological needs: relatedness. Values are on a 1--7 scale. Where all seven Advisees scored the scale floor, the bootstrap interval is degenerate ([1.00, 1.00]) and reflects a floor effect rather than high precision.}
\label{tab:relatedness}
\footnotesize\setlength{\tabcolsep}{4pt}
\begin{tabular}{lrlrl}
\toprule
\textbf{Advisor} & \textbf{Support} & \textbf{Support 95\% CI} & \textbf{Frustration} & \textbf{Frustration 95\% CI} \\
\midrule
GPT-4.1 & 4.52 & [4.00, 5.00] & 1.0 & [1.00, 1.00] \\
Claude-Sonnet-4.5 & 5.05 & [4.67, 5.38] & 1.0 & [1.00, 1.00] \\
Mistral-Medium & 3.86 & [3.38, 4.43] & 1.14 & [1.00, 1.33] \\
DeepSeek-V3.2 & 5.1 & [4.76, 5.43] & 1.0 & [1.00, 1.00] \\
\bottomrule
\end{tabular}
\end{table}

\begin{table}[H]
\centering
\caption{Basic psychological needs: overall net support, computed as support minus frustration averaged across the three dimensions. Larger values indicate better support of basic psychological needs.}
\label{tab:netsupport}
\footnotesize\setlength{\tabcolsep}{4pt}
\begin{tabular}{lr}
\toprule
\textbf{Advisor} & \textbf{Overall net support} \\
\midrule
GPT-4.1 & 3.52 \\
Claude-Sonnet-4.5 & 4.14 \\
Mistral-Medium & 1.73 \\
DeepSeek-V3.2 & 4.16 \\
\bottomrule
\end{tabular}
\end{table}

DeepSeek-V3.2 shows the largest support and lowest frustration on autonomy and relatedness; Claude-Sonnet-4.5 shows the largest support on competence. Mistral-Medium provides the lowest support and generates the highest frustration on all three dimensions. On overall net support DeepSeek-V3.2 ranks first (4.16) with Claude-Sonnet-4.5 a close second (4.14).

\subsection{Study 4 --- Support quality}

\noindent\textbf{Advisee} and \textbf{Advisor} as in Study 2. \textbf{Auditor:} gpt-5.4, applying three instruments with deliberately different access windows. CEISS\cite{munoz2025} sees only the Advisor's turns, so that crisis-intervention skill is rated independently of how the Advisee responded; MITI\cite{moyers2014} and the behaviour counter see the full history, since both code Advisor conduct relative to what the Advisee has just disclosed.

\begin{figure}[H]
\centering
\includegraphics[width=\textwidth]{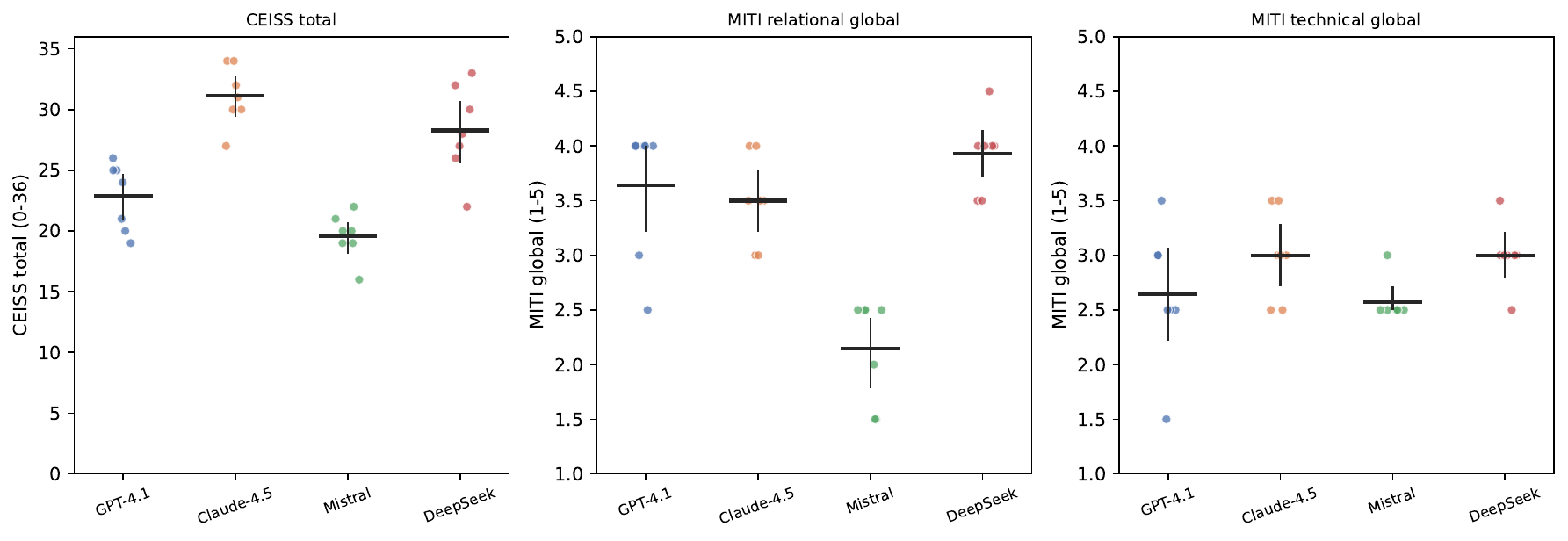}
\caption{Advisor performance on CEISS total (0--36) and the MITI global scores. Each point is one of the seven Advisees; error bars are 95\% percentile-bootstrap confidence intervals.}
\label{fig:quality}
\end{figure}

\begin{table}[H]
\centering
\caption{Support quality: CEISS. Subscale means are per-Advisee averages (items 0--4); the total 95\% CI is a percentile-bootstrap interval over the seven Advisees ($B = 10{,}000$), used because the scale is bounded and normality is not assumed.}
\label{tab:ceiss}
\resizebox{\textwidth}{!}{\footnotesize\setlength{\tabcolsep}{4pt}
\begin{tabular}{lrrrrrrrrrrr}
\toprule
\textbf{Advisor} & \textbf{C1} & \textbf{C2} & \textbf{C3} & \textbf{C4} & \textbf{C5} & \textbf{C6} & \textbf{C7} & \textbf{C8} & \textbf{C9} & \textbf{CEISS\_total} & \textbf{CEISS\_total 95\% CI} \\
\midrule
GPT-4.1 & 3.29 & 3.29 & 3.43 & 1.14 & 2.43 & 1.0 & 3.0 & 3.0 & 2.29 & 22.86 & [20.86, 24.71] \\
Claude-Sonnet-4.5 & 3.29 & 3.86 & 3.57 & 3.14 & 3.43 & 2.71 & 3.43 & 4.0 & 3.71 & 31.14 & [29.43, 32.71] \\
Mistral-Medium & 1.86 & 2.57 & 3.0 & 0.71 & 2.0 & 1.29 & 2.71 & 2.86 & 2.57 & 19.57 & [18.14, 20.71] \\
DeepSeek-V3.2 & 3.86 & 3.29 & 4.0 & 1.71 & 3.43 & 1.57 & 3.43 & 3.86 & 3.14 & 28.29 & [25.57, 30.71] \\
\bottomrule
\end{tabular}}
\end{table}

\begin{table}[H]
\centering
\caption{Support quality: MITI. The two ratio indices are reported as descriptive indicators only, with the number of conversations on which each is defined; see Methods for why percent complex reflection degenerates to a presence/absence indicator in this implementation. The relational and technical globals are the interpretable MITI measures here and are the ones entering the composite score.}
\label{tab:miti}
\footnotesize\setlength{\tabcolsep}{3.5pt}
\begin{tabular}{lcccccc}
\toprule
\textbf{Advisor} & \textbf{Relational global} & \textbf{Technical global} & \textbf{\% complex refl.} & \textbf{\emph{n} def.} & \textbf{Refl.:question} & \textbf{\emph{n} def.} \\
\midrule
GPT-4.1 & 3.64 [3.21, 4.00] & 2.64 [2.21, 3.07] & 1.0 [1.00, 1.00] & 6/7 & 1.79 [0.70, 2.95] & 7/7 \\
Claude-Sonnet-4.5 & 3.5 [3.21, 3.79] & 3.0 [2.71, 3.29] & 1.0 [1.00, 1.00] & 5/7 & 0.27 [0.13, 0.40] & 7/7 \\
Mistral-Medium & 2.14 [1.79, 2.43] & 2.57 [2.50, 2.71] & 0.93 [0.80, 1.00] & 5/7 & 1.0 [0.00, 2.00] & 2/7 \\
DeepSeek-V3.2 & 3.93 [3.71, 4.14] & 3.0 [2.79, 3.21] & 1.0 [1.00, 1.00] & 6/7 & 0.85 [0.43, 1.30] & 7/7 \\
\bottomrule
\end{tabular}
\end{table}

\begin{figure}[H]
\centering
\includegraphics[width=\textwidth]{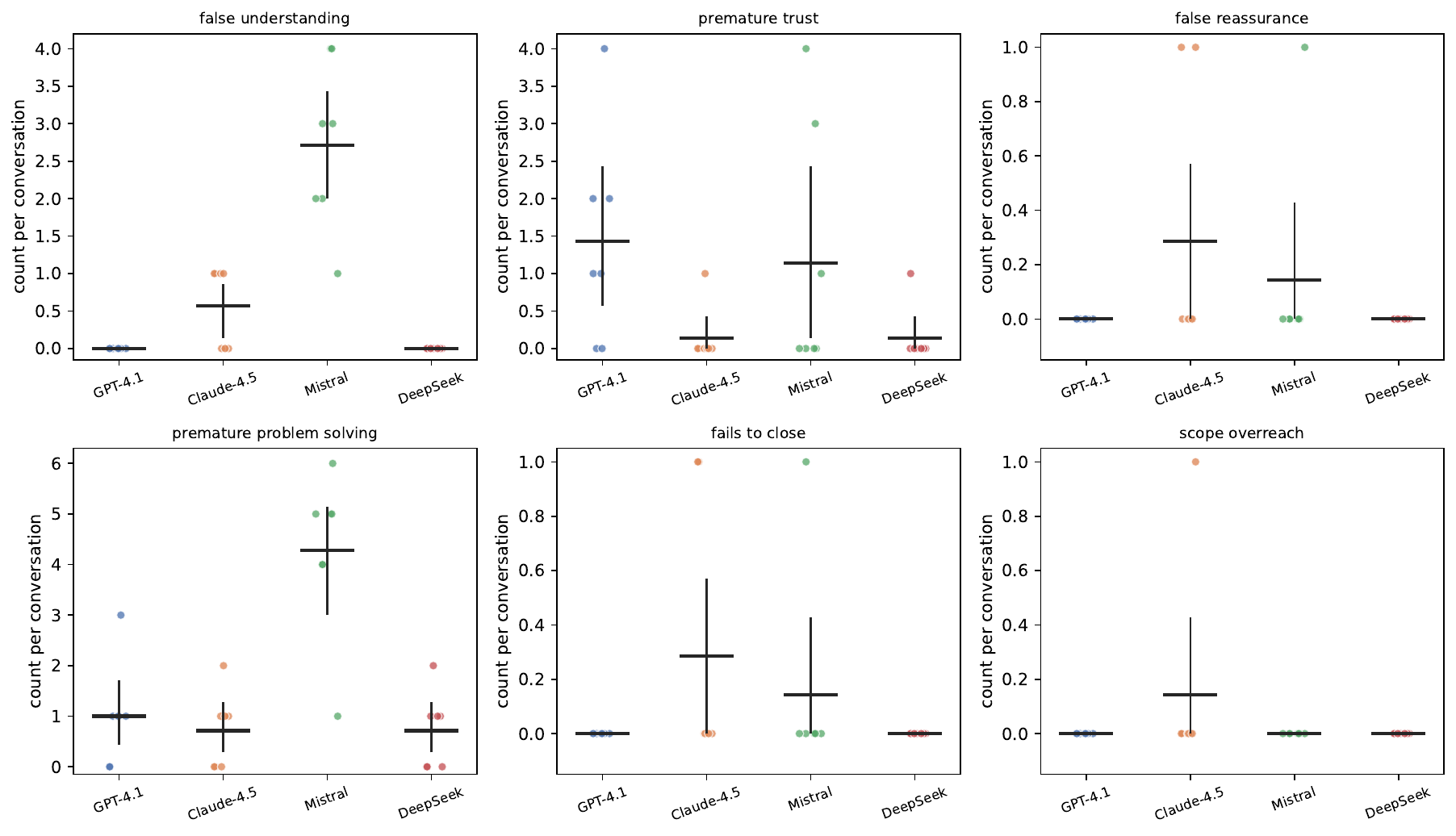}
\caption{Problematic Advisor behaviours, counted per conversation. Each point is one of the seven Advisees; error bars are 95\% percentile-bootstrap confidence intervals. A larger value means more instances of problematic behaviour.}
\label{fig:behaviour}
\end{figure}

\begin{table}[H]
\centering
\caption{Behavioural counter: reactance-producing expressions. Counts are per conversation, averaged across the seven Advisees, with 95\% percentile-bootstrap confidence intervals. Behaviours never observed for an Advisor give a degenerate interval at zero, a true absence rather than high precision.}
\label{tab:behreact}
\small\setlength{\tabcolsep}{5pt}
\begin{tabular}{lccc}
\toprule
\textbf{Advisor} & \textbf{False understanding} & \textbf{Premature trust} & \textbf{False reassurance} \\
\midrule
GPT-4.1 & 0.0 [0.00, 0.00] & 1.43 [0.57, 2.43] & 0.0 [0.00, 0.00] \\
Claude-Sonnet-4.5 & 0.57 [0.14, 0.86] & 0.14 [0.00, 0.43] & 0.29 [0.00, 0.57] \\
Mistral-Medium & 2.71 [2.00, 3.43] & 1.14 [0.14, 2.43] & 0.14 [0.00, 0.43] \\
DeepSeek-V3.2 & 0.0 [0.00, 0.00] & 0.14 [0.00, 0.43] & 0.0 [0.00, 0.00] \\
\bottomrule
\end{tabular}
\end{table}

\begin{table}[H]
\centering
\caption{Behavioural counter: process failures.}
\label{tab:behproc}
\small\setlength{\tabcolsep}{5pt}
\begin{tabular}{lcc}
\toprule
\textbf{Advisor} & \textbf{Premature problem-solving} & \textbf{Rushing past emotion} \\
\midrule
GPT-4.1 & 1.0 [0.43, 1.71] & 0.0 [0.00, 0.00] \\
Claude-Sonnet-4.5 & 0.71 [0.29, 1.29] & 0.0 [0.00, 0.00] \\
Mistral-Medium & 4.29 [3.00, 5.14] & 0.0 [0.00, 0.00] \\
DeepSeek-V3.2 & 0.71 [0.29, 1.29] & 0.0 [0.00, 0.00] \\
\bottomrule
\end{tabular}
\end{table}

\begin{table}[H]
\centering
\caption{Behavioural counter: boundary and scope.}
\label{tab:behbound}
\small\setlength{\tabcolsep}{5pt}
\begin{tabular}{lcc}
\toprule
\textbf{Advisor} & \textbf{Fails to close} & \textbf{Scope overreach} \\
\midrule
GPT-4.1 & 0.0 [0.00, 0.00] & 0.0 [0.00, 0.00] \\
Claude-Sonnet-4.5 & 0.29 [0.00, 0.57] & 0.14 [0.00, 0.43] \\
Mistral-Medium & 0.14 [0.00, 0.43] & 0.0 [0.00, 0.00] \\
DeepSeek-V3.2 & 0.0 [0.00, 0.00] & 0.0 [0.00, 0.00] \\
\bottomrule
\end{tabular}
\end{table}

\begin{table}[H]
\centering
\caption{Behavioural counter: overall indices.}
\label{tab:behoverall}
\footnotesize\setlength{\tabcolsep}{3.5pt}
\begin{tabular}{lccccc}
\toprule
\textbf{Advisor} & \textbf{Reactance total} & \textbf{Boundary overstep} & \textbf{Talk-listen ratio} & \textbf{Advisor words (mean)} & \textbf{Long turns} \\
\midrule
GPT-4.1 & 1.43 [0.57, 2.43] & 0.0 [0.00, 0.00] & 1.84 [1.50, 2.16] & 312.95 [244.32, 382.48] & 4.71 [4.00, 5.43] \\
Claude-Sonnet-4.5 & 1.0 [0.57, 1.43] & 0.43 [0.14, 0.86] & 1.88 [1.45, 2.40] & 354.41 [278.87, 444.05] & 4.57 [4.14, 5.14] \\
Mistral-Medium & 4.0 [2.57, 5.57] & 0.14 [0.00, 0.43] & 3.26 [2.69, 3.84] & 586.63 [495.68, 687.66] & 4.71 [4.29, 5.29] \\
DeepSeek-V3.2 & 0.14 [0.00, 0.43] & 0.0 [0.00, 0.00] & 1.79 [1.24, 2.41] & 276.27 [198.35, 368.55] & 3.29 [2.57, 4.00] \\
\bottomrule
\end{tabular}
\end{table}

\subsection{Between-Advisor comparisons}

\begin{table}[H]
\centering
\footnotesize
\caption{Between-Advisor comparisons, blocking on Advisee. One-way repeated-measures analyses with Advisee as the random factor, since the same seven Advisees appear in all four arms. Friedman's rank test accompanies each contrast as a distribution-free check. Holm correction is applied within three declared families: TENS subscales, support-quality metrics and behaviour counts. Advisee variance is the proportion of total variance attributable to the Advisee rather than the Advisor. Rushing past emotion was constant at zero across the corpus and is not testable.}
\label{tab:blocked}
\resizebox{\textwidth}{!}{\footnotesize\setlength{\tabcolsep}{4pt}
\begin{tabular}{llrlrrrrrl}
\toprule
\textbf{Family} & \textbf{Measure} & \textbf{F} & \textbf{df} & \textbf{p} & \textbf{partial eta\^{}2} & \textbf{Friedman p} & \textbf{Advisee variance} & \textbf{Holm p (within family)} & \textbf{Advisors differ} \\
\midrule
TENS subscales & autonomy\_support & 5.66 & 3,18 & 0.0065 & 0.485 & 0.0184 & 23\% & 0.013 & yes \\
TENS subscales & autonomy\_frustration & 16.67 & 3,18 & 0.0 & 0.735 & 0.0004 & 12\% & 0.0001 & yes \\
TENS subscales & competence\_support & 7.88 & 3,18 & 0.0015 & 0.568 & 0.0079 & 23\% & 0.0044 & yes \\
TENS subscales & competence\_frustration & 12.5 & 3,18 & 0.0001 & 0.676 & 0.0013 & 13\% & 0.0006 & yes \\
TENS subscales & relatedness\_support & 10.13 & 3,18 & 0.0004 & 0.628 & 0.0054 & 35\% & 0.0016 & yes \\
TENS subscales & relatedness\_frustration & 2.08 & 3,18 & 0.1391 & 0.257 & 0.1116 & 20\% & 0.1391 & no \\
TENS subscales & net support & 13.61 & 3,18 & 0.0001 & 0.694 & 0.0006 & 17\% & 0.0004 & yes \\
Support quality & CEISS total & 30.61 & 3,18 & 0.0 & 0.836 & 0.0005 & 10\% & 0.0 & yes \\
Support quality & MITI relational global & 21.49 & 3,18 & 0.0 & 0.782 & 0.0016 & 9\% & 0.0 & yes \\
Support quality & MITI technical global & 2.22 & 3,18 & 0.1206 & 0.27 & 0.0905 & 21\% & 0.1206 & no \\
Behaviour counts & false\_understanding & 36.26 & 3,18 & 0.0 & 0.858 & 0.0003 & 8\% & 0.0 & yes \\
Behaviour counts & premature\_trust & 2.35 & 3,18 & 0.1065 & 0.281 & 0.0786 & 16\% & 0.5078 & no \\
Behaviour counts & false\_reassurance & 1.74 & 3,18 & 0.1953 & 0.224 & 0.194 & 35\% & 0.5859 & no \\
Behaviour counts & premature\_problem\_solving & 19.71 & 3,18 & 0.0 & 0.767 & 0.0027 & 10\% & 0.0 & yes \\
Behaviour counts & rushing\_emotion & --- & --- & --- & --- & --- & --- & --- & --- \\
Behaviour counts & fails\_to\_close & 1.27 & 3,18 & 0.3148 & 0.175 & 0.2998 & 16\% & 0.6296 & no \\
Behaviour counts & scope\_overreach & 1.0 & 3,18 & 0.4155 & 0.143 & 0.3916 & 22\% & 0.6296 & no \\
Behaviour counts & reactance\_total & 11.15 & 3,18 & 0.0002 & 0.65 & 0.0008 & 11\% & 0.0014 & yes \\
Behaviour counts & boundary\_overstep\_total & 2.4 & 3,18 & 0.1016 & 0.286 & 0.1116 & 12\% & 0.5078 & no \\
\bottomrule
\end{tabular}}
\end{table}

\subsection{Overall results and ranking}

Each dimension is scaled 0--100 against its instrument's theoretical range, so that higher is always better. Emotion handling is within-session change as a fraction of achievable improvement: PSS $(\text{start}-\text{end})/\text{start}$, PANAS-positive $(\text{end}-\text{start})/(5-\text{start})$, PANAS-negative $(\text{start}-\text{end})/(\text{start}-1)$. Basic psychological needs uses TENS-Life (1--7) with support scaled $(x-1)/6$ and frustration reverse-scored $(7-x)/6$. Intervention quality is the equal-weight mean of $\text{CEISS}_{\text{total}}/36$, $(\text{relational global}-1)/4$ and $(\text{technical global}-1)/4$; reflective listening enters once, through the relational global, and percent complex reflection and the reflection-to-question ratio are reported as indicators but excluded from the composite to avoid triple-counting. Behaviour (conduct) is $1 - (\text{problematic behaviours per Advisor turn})/2$, from the behaviour counter. The composite is the unweighted mean of all four dimensions, computed from the unrounded dimension means.

\begin{figure}[H]
\centering
\includegraphics[width=\textwidth]{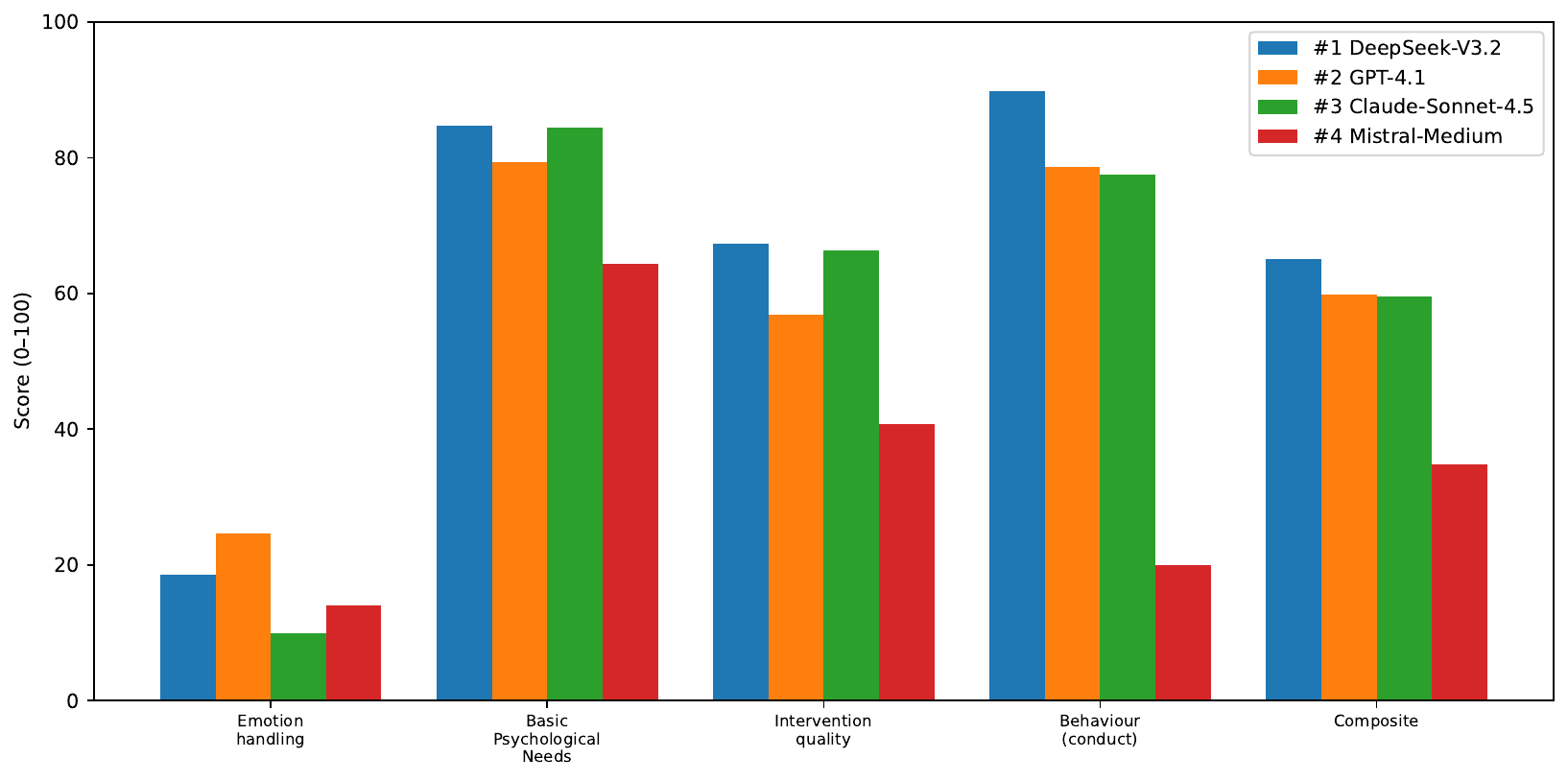}
\caption{Overall results from Studies 2, 3 and 4, and the composite score. Confidence intervals for the composite are given in Table~\ref{tab:composite}.}
\label{fig:rankings}
\end{figure}

\begin{table}[H]
\centering
\caption{Overall Advisor ranking with confidence intervals. Intervals are percentile bootstrap over the seven Advisees ($B = 10{,}000$).}
\label{tab:composite}
\footnotesize\setlength{\tabcolsep}{4pt}
\begin{tabular}{rllc}
\toprule
\textbf{Rank} & \textbf{Advisor} & \textbf{Composite} & \textbf{95\% CI} \\
\midrule
1 & DeepSeek-V3.2 & 65.0 & [62.2, 68.2] \\
2 & GPT-4.1 & 59.9 & [52.2, 64.8] \\
3 & Claude-Sonnet-4.5 & 59.6 & [55.9, 63.6] \\
4 & Mistral-Medium & 34.8 & [28.2, 41.6] \\
\bottomrule
\end{tabular}
\end{table}

\begin{table}[H]
\centering
\caption{Pairwise differences in composite score, using the same resampled Advisee draws for both members of each pair.}
\label{tab:comppair}
\footnotesize\setlength{\tabcolsep}{4pt}
\begin{tabular}{lrlrl}
\toprule
\textbf{Comparison} & \textbf{Difference} & \textbf{95\% CI} & \textbf{p (boot)} & \textbf{CI excludes 0} \\
\midrule
GPT-4.1 - Claude-Sonnet-4.5 & 0.3 & [-7.2, 6.5] & 0.869 & no \\
GPT-4.1 - Mistral-Medium & 25.1 & [18.9, 31.5] & 0.0 & yes \\
GPT-4.1 - DeepSeek-V3.2 & -5.2 & [-11.7, -0.3] & 0.031 & yes \\
Claude-Sonnet-4.5 - Mistral-Medium & 24.7 & [16.2, 32.2] & 0.0 & yes \\
Claude-Sonnet-4.5 - DeepSeek-V3.2 & -5.5 & [-10.9, -0.5] & 0.034 & yes \\
Mistral-Medium - DeepSeek-V3.2 & -30.2 & [-34.2, -26.2] & 0.0 & yes \\
\bottomrule
\end{tabular}
\end{table}

\section{Discussion}

We introduce a framework for studying human--AI conversations that includes Advisors (for example, a therapist), Advisees (for example, a help-seeker) and Auditors (who assess the quality of the conversation). Each of the three can be either an AI or a human. The framework is used here in the context of understanding the quality of mental health crisis conversations in a psychological-profile-informed way across four studies.

The most consequential result is not how the four models ordered but what they shared: the framework exposed the same failure modes in every one of them. All four Advisors talked more than they listened, with Advisor-to-Advisee word ratios above one in every case (1.79--3.26) and mean Advisor turns of roughly 280 to 590 words, which implies a weak sense of language economy that poorly matches the low attentional budget of someone in acute distress. As noted in the Introduction, acute stress of this kind impairs prefrontal executive function and narrows attention toward high-priority information, reducing the working-memory resources available to process new input; lengthy Advisor turns are therefore especially costly in this state\cite{shields2016}. All four also tended to move to problem-solving before the Advisee's situation had been explored, most severely in Mistral-Medium. Two predicted problems, by contrast, largely failed to materialise in this scenario: the behaviour counter recorded almost no instances of rushing past an emotional disclosure, and boundary or scope overreach was rare.

The trait-recovery experiment (Study 1) showed that specified psychological profiles are expressed in the generated dialogue strongly enough for an independent Auditor, blind to the specification, to reconstruct them: band-score correlations reached $r$ = 0.78. This means that the Advisor evaluation can be made personality-aware. The placebo control establishes that this is a property of the enacted profile rather than of the shared scenario or biography: with the specification removed and demographics held byte-identical, recovery fell by roughly half in every model, and every $\Delta r$ interval excluded zero.

Recovery held on every instrument. With the four-Auditor ensemble, agreement was excellent for the NEO-FFI dimensions (ICC(2,4) = .964) and good for the constructs that lie outside the lexical tradition: resilience (.870), coping style (.827), coping self-efficacy (.817) and reactance (.791). This was not the expected outcome. These are appraisals assessed almost exclusively by questionnaire, and we are not aware of prior evidence that a person's standing on them can be read off conversational language at all; the anticipation set out in the Introduction was that they would be recovered considerably less well than the lexically derived dimensions, if at all. That an ensemble of specification-blind Auditors recovers them at this level is the more surprising half of the result.

A gradient does appear, but only when the ensemble is dissolved. For a single Auditor, agreement runs from .871 on the NEO-FFI dimensions down to .626 for resilience, .545 for coping style, .527 for coping self-efficacy and .486 for reactance, and that ordering reproduces one of the more robust findings in person perception: judges agree more about dispositions expressed in observable conduct than about those that are internal or evaluative\cite{john1993,ashton2025}. The same prediction follows from the self--other knowledge asymmetry model, in which an observer has access to conduct but not to appraisal\cite{vazire2010}, and from meta-analytic work on observer ratings across the Big Five\cite{connelly2010}. What the ensemble does is compensate for exactly this: aggregating four independent judges lifts every construct into a usable range, which is why the panel rather than the individual Auditor is the unit of the framework. The practical implication follows directly: personality-aware auditing of internal appraisals requires an ensemble, whereas dispositions that surface in speech can be scored more economically.

That the Big Five dimensions are the ones recovered most reliably is not an artefact of having specified them, and the manner of specification is what makes the result informative. The Advisee prompt supplied no trait-descriptive vocabulary. It fixed a band on each dimension of a psychometrically validated instrument, defined by score intervals rather than by adjectives, and said nothing about how that band should be expressed in words. Recovering it therefore required two independent steps to hold. The Advisee model had to translate a quantitative coordinate into sustained conversational conduct, and a specification-blind Auditor had to invert that mapping from the transcript alone. Band-score correlations of $r$ = 0.78, with ICC(2,4) = .964 across the NEO-FFI dimensions, indicate that both steps held.

This is more than the lexical tradition on its own would guarantee. The Big Five rest on the lexical hypothesis, that socially consequential individual differences come to be encoded in natural language, and were themselves arrived at by factor-analysing that vocabulary\cite{goldberg1990}, and standing on these dimensions has been recovered from spontaneous text across diaries, essays, correspondence and social media consistently enough for language-based assessment to converge with both self-reports and informant reports\cite{park2015}. Language is therefore the medium in which Big Five content is expected to surface, rather than an impoverished substitute for watching someone behave. What that literature establishes runs from language to estimate, in speakers whose position was never prescribed. The direction demonstrated here is the reverse and under specification: a position on a validated questionnaire is written into a generated interlocutor and read back out of its dialogue by an independent judge. That round trip is what licenses treating a synthetic Advisee as an instance of a defined profile rather than as an undifferentiated conversational partner, and it is the condition on which personality-aware evaluation of Advisors depends. Two things separate it from existing psychometric work on synthetic personality in LLMs\cite{serapiogarcia2025}. The verification route differs: rather than administering the inventory back to the model, the prescribed profile is recovered by an independent judge from dialogue the model produced with a third party, so what is tested is expression in interaction rather than self-report. And the coverage differs: the specification is not confined to the Five Factor Model but extends to coping style, coping self-efficacy, resilience and reactance. These are general dispositions, motivational and regulatory rather than descriptive, that no taxonomy of trait adjectives was built to represent; the crisis scenario is where we test them, not the limit of where they apply.

What separates the Advisors is not how the Advisee ended up feeling but what the Advisor did. All four moved all three emotion measures in the predicted direction, yet an omnibus test blocking on Advisee detected no reliable difference between them on any outcome (perceived stress $F(3,18) = 2.36$, $p = .106$; positive affect $F(3,18) = 0.95$, $p = .438$; negative affect $F(3,18) = 2.11$, $p = .135$), and the Advisee accounted for 45--56\% of the variance in emotional change, more than the Advisor did. The process measures behave differently. Blocking on Advisee and correcting within families, the Advisors differed on crisis-intervention skill (CEISS total $F(3,18) = 30.61$, Holm $p < .001$, partial $\eta^2$ = .84), on the motivational-interviewing relational global ($F(3,18) = 21.49$, Holm $p < .001$), on net need support ($F(3,18) = 13.61$, Holm $p < .001$) and on five of its six subscales, and on three behaviour counts: false understanding ($F(3,18) = 36.26$, Holm $p < .001$), premature problem-solving ($F(3,18) = 19.71$, Holm $p < .001$) and total reactance-producing expressions ($F(3,18) = 11.15$, Holm $p = .001$). The MITI technical global, relatedness frustration and the remaining behaviour counts did not differ. On the composite score, DeepSeek-V3.2 led at 65.0 [62.2, 68.2] ahead of GPT-4.1 at 59.9 [52.2, 64.8] and Claude-Sonnet-4.5 at 59.6 [55.9, 63.6], which were themselves indistinguishable (difference 0.3 [$-$7.2, 6.5], $p$ = .87); Mistral-Medium was last by a wide margin at 34.8 [28.2, 41.6]. DeepSeek-V3.2's lead over the two middle models is narrow but reliable (5.2 [0.3, 11.7], $p$ = .031; 5.5 [0.5, 10.9], $p$ = .034). Outcome measures do not discriminate among these Advisors at this sample size; conduct measures do.

Using this framework, we find that synthetic red-teaming approaches can be useful to compare the quality of Advisor agents provided by different companies in the context of mental health crisis support. We compared four widely available models from different providers; any provider-level differences observed here confound model architecture, training data and safety policies, and should not be attributed to a single factor. The results reinforce the case that response-level benchmarks, including mental-health-specific ones, are necessary but not sufficient for conversational clinical agents: the differences we observed emerged only over multi-turn interaction under sustained distress. We also find that the framework is useful to provide evidence that can be used to improve agents and to make procurement decisions for organisations that build on these models.

A further limitation concerns the Auditor in Studies 2 to 4. Where Study 1 used an ensemble of four models and reports inter-Auditor agreement, Studies 2 to 4 rely on a single Auditor, gpt-5.4, and no inter-Auditor agreement is available for them. Because gpt-5.4 shares a provider with one of the four Advisors it scores, the between-Advisor differences reported above are open to a family-preference objection that the present design cannot rule out. Two observations bear on it, both weak. The single Auditor did not place its own provider's model first: GPT-4.1 ranks second on the composite, behind DeepSeek-V3.2 and within the confidence interval of Claude-Sonnet-4.5. And in Study 1, where the cross-tabulation of Auditor against instantiating model can be examined directly, self-preference was not a uniform property of the judges. Neither observation substitutes for an ensemble, and replication of Studies 2 to 4 with a panel of Auditors is the first item of further work.

Several further limitations should be borne in mind. First, the scenario was deliberately held constant: every synthetic Advisee was a family caregiver who had just learned of a relative's dementia diagnosis, situated in a British setting. Fixing the scenario is what allows the framework to isolate how Advisor quality varies across psychological profiles, but it also means that the present findings should be read as a proof of concept rather than as evidence that generalises across crisis types, clinical conditions or cultural contexts. Second, the analysis rests on seven profiles, with Advisors at temperature zero and Advisees at 0.7, without within-cell replication; all inferential statistics are therefore exploratory. In particular, the absence of significant emotion-stabilisation effects for most of the twelve contrasts after correction for multiplicity is almost certainly a matter of statistical power rather than evidence of no effect --- the PSS deltas for Mistral-Medium ($-$0.16) and DeepSeek-V3.2 ($-$0.20) run in the same direction and are comparable in magnitude to that of GPT-4.1 ($-$0.31) --- so counting the number of significant tests is not a sound index of Advisor quality, and effect sizes with confidence intervals should be preferred. Third, the Auditor scores are inferences drawn from the conversation text by a language model; they should not be equated with measured emotional states of a human, and human clinical experts remain necessary in the loop. Framed this way, the ceiling effects we observed are informative precisely because they are not an artefact of a single model but recur across Auditors.

These limitations map onto clear directions for future work. The most immediate is to vary the scenario systematically, across crisis types, clinical conditions and, in particular, cultural contexts, since language models are known to carry a Western cultural bias that could distort support delivered to caregivers from other backgrounds\cite{moore2025,tao2024}. A second priority is to complement the Auditor with a human-validation sub-study, in which clinical psychologists rate a stratified subsample of conversations, so that claims about the representativeness of the synthetic Advisees can be defended empirically. Third, the two matched pairs specified in the design --- one isolating gender and one isolating age, with every other psychometric and demographic variable held constant --- support controlled analyses of whether Advisor quality varies with these characteristics, which we intend to report once the corresponding data are available. Finally, enlarging the cohort and introducing within-cell replication would move the framework from the present exploratory footing toward confirmatory inference.

People reach for LLMs precisely at moments of distress, under high uncertainty, low controllability and intense emotion, yet these are the moments in which a person's capacity to take in new information is most reduced. Across four studies and every model tested, the framework exposed a shared and consequential mismatch with that constraint: the Advisors talked more than they listened and moved to problem-solving before the caregiver's situation had been explored, running counter to the listen-before-solving sequence that evidence-based crisis support requires\cite{roberts2005,wang2024}. This tendency is not a peripheral stylistic quirk but a substantive limitation for crisis use, because verbosity imposes exactly the kind of processing load that acute stress leaves least available. Encouragingly, the failures were of proportion rather than of catastrophe --- the more dangerous behaviours we anticipated seldom appeared --- which suggests that calibrating Advisors toward listening, brevity and stabilisation is a tractable and high-value target. Synthetic, profile-aware red teaming of the kind demonstrated here offers a practical route to that calibration and to the procurement and safety decisions that will shape how these systems support people in crisis.

\section{Methods}

We identified three roles for this study, all of which can be human or synthetic. Advisees are those with multiple psychological profiles that receive health advice; these can be help-seekers, or simulated Advisees, for example in training scenarios. Advisors are the different agents which provide advice, for example a human doctor or a mental health chatbot. Auditors are the agents used to assess the quality of the Advisors.

Identifying these roles and separating them from the embodiment, that is, human or synthetic, or different foundational models, allows for systems thinking and clearer abstractions that can be implemented in software platforms. For example, a software Auditor may be used both for checking on the safety and efficacy of human--AI support conversations, where the human is a help-seeker and the AI an Advisor, or the other way around, as in simulated patient scenarios, and also for equivalent human--human conversations. The outcome measures used to assess human--human therapies can more readily be used to improve human--AI ones.

\subsection{The Advisee--Advisor--Auditor (AAA) framework}

To assess risks in the advice provided by Advisors, we can use synthetic patients. The process of evaluating how good the advice given to someone in distress is can be mapped to the interaction of the triad Advisor--Advisee--Auditor. The Advisee is the one seeking help or advice and can be a synthetic patient or a real human.

We use instances of acute mental health problems as context for our study. We focus on acute crisis because it is both the context in which conversational support is most needed and the one in which it is most demanding; it is also where documented LLM failures carry the greatest risk\cite{moore2025}. Methodologically, this context is well suited to our aim: the precipitating event can be held constant across Advisees, so that variation in Advisor behaviour can be attributed to the psychological profile rather than to the situation. Specifically, our case is a situation where an Advisee receives bad news about the health of a close relative. In that situation, specific psychological profiles are known to predict caregiver behaviour and are plausibly associated with the specific challenges that the interaction poses to AI and human Advisors: coping strategies as measured by the Coping with Health Injuries and Problems inventory (CHIP)\cite{endler1998}, traits in the Five Factor Theory of Personality as measured by the NEO-FFI\cite{costa1992}, reactance as measured by the Hong Psychological Reactance Scale\cite{hong1996}, coping self-efficacy as measured by the CSES-8\cite{chesney2006}, and resilience and strain in caregivers as measured by ResQ-Care\cite{resqcare}. The associations between these dimensions and caregiver outcomes are small but consistent\cite{bucher2019}; whether Advisor quality is in fact moderated by profile is an open question that the present design can raise but not settle.

\subsection{Advisee specification}

Throughout this paper, a personality profile denotes a person's standing on the five dispositional dimensions specified here, and not standing on the Five Factor Model alone. Three of those dimensions are FFM traits, but coping style, coping self-efficacy, resilience and reactance proneness are equally dispositional: they are characteristics of a person rather than features of a situation, and they are stable enough to be prescribed in advance and recovered afterwards. What distinguishes them from the FFM dimensions is not that they are less trait-like but that they are motivational, regulatory and self-appraisal constructs, which the lexical taxonomies that produced the Five Factor Model were never built to represent.

The five dimensions were not chosen arbitrarily. We conducted a review of the literature on the psychological variables that most consistently predict the mental-health trajectory (depression, anxiety and perceived burden) of family caregivers of people with Alzheimer's disease and related dementias, in order to specify profiles that are most critical and most likely to elicit difficult responses in moments of crisis and bad news. Trait neuroticism predicts higher depression risk, greater objective and subjective burden, and a poorer response to caregiver interventions\cite{melo2011,jang2004}. Coping style is longitudinally associated with psychological morbidity, with dysfunctional coping predicting worse anxiety and depression and emotion-focused coping showing protective effects\cite{delpino2011,cooper2008}. Caregiving self-efficacy buffers the impact of stressors on depression and burden, acting as a protective resource\cite{gallagher2011,rabinowitz2009}. Psychological reactance is particularly relevant in an advice-giving context, since a controlling communication style threatens perceived autonomy and predicts rejection of the advice offered\cite{rains2013,dillard2005}. Resilience operates as a protective factor, being inversely associated with depression, anxiety and burden in dementia caregivers\cite{poe2023}. Selecting profiles that span these five predictors therefore yields synthetic Advisees that are both clinically plausible and informative for stress-testing crisis-support behaviour.

Four of the seven Advisees form two matched pairs, identical on every psychometric and demographic dimension except for a single manipulated variable: one pair differs only in gender and the other only in age (59 versus 91 years). The design therefore incorporates a gender-controlled and an age-controlled comparison rather than a post-hoc observation. We specified the psychological profiles of seven synthetic Advisees, for which we chose a three-level scale: low, medium and high for each trait.

This allowed us to repurpose the questionnaires for prompting the Advisees. For example, if an Advisee is expected to score high on the palliative coping strategy according to the CHIP questionnaire, we add an instruction stating that an item concerning remaining in bed is something the Advisee is likely to say or agree with when reacting to health problems such as illnesses, sickness and injuries. Conversely, if the Advisee is expected to score low on the palliative coping strategy, a prompt instruction would state that an item concerning resting when tired is something the Advisee is unlikely to say or identify with in that situation. We do not give sample phrases when the levels are set to medium. As ways to generate diversity in the backstory and grounding for the simulated interactions we also specify some biographical and demographic information in the form of key--value pairs, which is also passed to the system prompt.

\subsection{The no-profile baseline}

To provide a placebo control for Study 1, each of the seven Advisees was additionally instantiated with all five psychometric specification blocks removed from the system prompt --- the CHIP coping specification with its representative phrases, the NEO-FFI levels, CSES-8, Hong reactance and ResQ-Care --- together with the instruction to enact the profile and the word ``personality'' from the list of biographical attributes the Advisee is told to draw on. This reduced the Advisee prompt from 6,243 to 1,723 characters. All demographic and biographical attributes were verified programmatically to be byte-identical between the two arms --- country, region, age, gender, ethnicity, education, occupation, income band, household composition, health status, caring responsibilities, internet access, technology familiarity and device ownership --- as were the scenario and task strings. The Advisor was GPT-4.1 at temperature 0 with the provider default minimal system prompt in all eight runs, and the Advisee temperature (0.7) and turn cap (10) were likewise identical; the only differences across configurations were the instantiating model and the system-prompt file. This yields 28 baseline conversations mirroring the 28 specified-profile conversations, 56 in total for Study 1. The manipulation removes both the trait specifications and the instruction to enact them; these cannot be varied independently, since an instruction to enact an unspecified profile is not well defined.

\begin{figure}[H]
\centering
\includegraphics[width=0.85\textwidth]{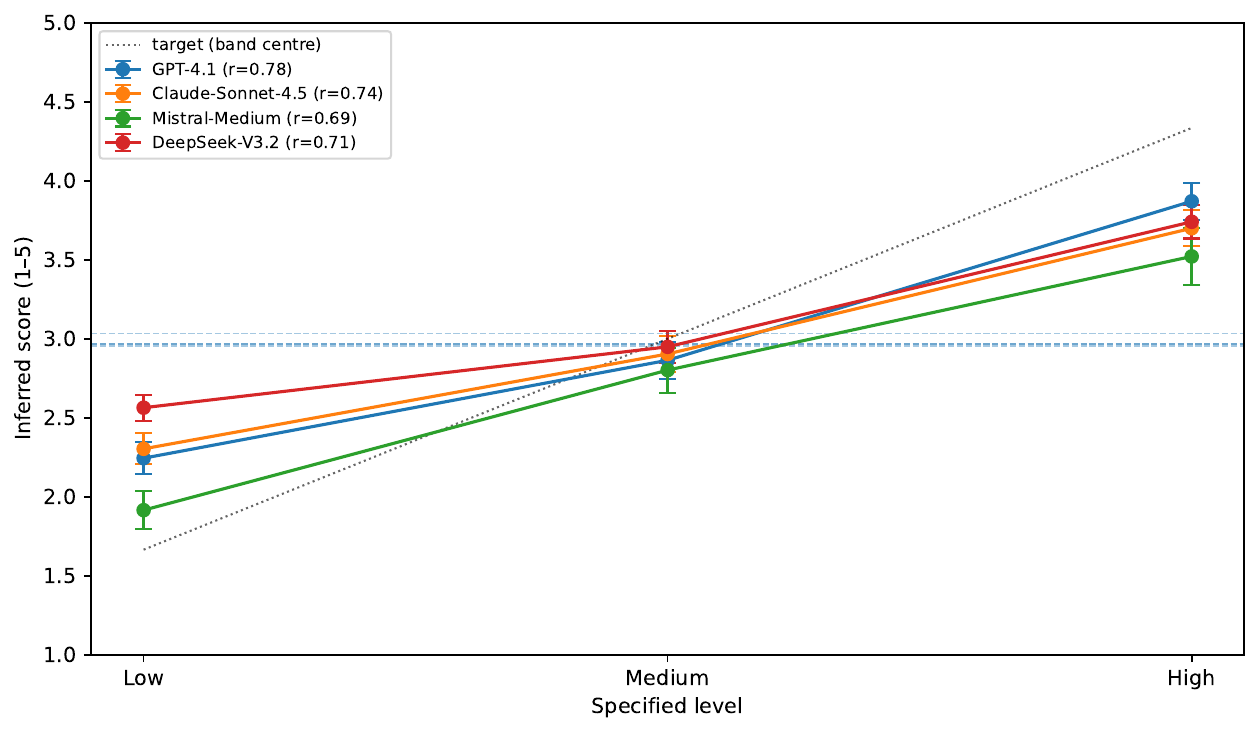}
\caption{Recovered score against specified level, by instantiating model. Points are mean recovered scores across the seven Advisees and ten constructs for each specified band, with 95\% confidence intervals. The dotted line is the target band centre. Recovery is compressed toward the scale midpoint at both extremes, which accounts for the gap between the band-score correlations ($r$ = 0.69--0.78) and in-band accuracy (50--63\%).}
\label{fig:enactment}
\end{figure}

\subsection{Advisor specification}

We aim to compare widely available models with minimal additional safeguarding, so that observed conduct is attributable to the model rather than to a provider's product layer. We chose among popular models as representative of what is available commercially and we keep the system prompt instructions very simple.

All models were accessed through Azure AI Foundry between June and August 2026, with Azure content filtering set to its minimum configuration. Advisor models were gpt-4.1 (version 2025-04-14), claude-sonnet-4-5 (version 20250929), mistral-medium-2505 (version 1) and DeepSeek-V3.2 (version 1), each run at temperature 0 with the provider default minimal system prompt. The same four models were used to instantiate Advisees, at temperature 0.7. Auditors in Study 1 were these four models used as an ensemble; Studies 2 to 4 used gpt-5.4 (version 2026-03-05). Conversations were capped at ten turns per participant and terminated on the token \texttt{/end}. Because the models were accessed with minimal additional safeguarding rather than through their consumer-facing applications, the conduct reported here should be attributed to the models themselves rather than to any product-layer moderation; commercially deployed assistants built on these models may behave differently.

\subsection{Auditor specification}

The Auditors are composed of multiple outcome assessments or questionnaires, which can be combined for different studies. These include the Motivation, Engagement and Thriving in User Experience scales (METUX)\cite{burnell2023,peters2018}, the Basic Psychological Need Satisfaction and Frustration Scale (TENS)\cite{chen2015}, the Crisis and Emergency Intervention Skills Scale (CEISS)\cite{munoz2025}, the Motivational Interviewing Treatment Integrity Code (MITI)\cite{moyers2014}, the Positive and Negative Affect Schedule (PANAS)\cite{watson1988}, and the Perceived Stress Scale (PSS)\cite{cohen1983}.

Each Auditor may have a different level of access to conversational turns. PANAS and PSS have access only to the first two and last two turns by the Advisee, to see the effect of the support interaction in the dialogue. Both instruments were administered in their state format, referring to how the Advisee feels at that point in the conversation; the PSS has been validated in both trait and state formats, and the state format is the one appropriate to change within a single session. TENS-Life and MITI have access to the full history including turns by both the Advisee and the Advisor. The CEISS Auditor has access only to turns by the Advisor. The trait battery in Study 1 has access to the full history including turns by both the Advisee and the Advisor; the instruction to judge only the Advisee constrains the target of the rating, not the evidence available for it, since the Advisor's turns provide the context in which the Advisee's reactions are interpreted.

Auditors received no profile prompt, no demographic block and no scenario briefing --- only the instrument item and the transcript. Advisees disclose some biographical detail in conversation, but identically in both conditions, since demographics were held constant; the Auditors were therefore blind to the psychometric specification, which is the manipulated variable. Baseline and specified-profile conversations were scored with an identical prompt containing no condition label, and each item was scored in an independent, stateless request, so the Auditor could not distinguish the two conditions except from the conversation itself.

Our MITI implementation assigns at most one MITI code per Advisor turn, whereas MITI 4.2.1 codes individual utterances. Because a turn containing any complex reflection is coded as such, no simple reflection was coded anywhere in the corpus, and percent complex reflection therefore degenerates to a presence/absence indicator rather than a proportion. The reflection-to-question ratio is undefined where the Advisor asked no questions. Both are reported as descriptive indicators only; the relational and technical globals are the interpretable MITI measures and are the ones entering the composite score.

\subsection{The behavioural counter}

Some assessments count instances of specific Advisor behaviours. We designed a behaviour-counter Auditor whose task is to identify Advisor turns exhibiting behaviours that may be acceptable, or even helpful, in ordinary conversation but are commonly regarded as counterproductive in an acute crisis context. The catalogue draws on three convergent traditions: crisis intervention theory, which frames the acute encounter as requiring stabilisation and rapport before problem-solving\cite{roberts2005,wang2024,hobfoll2007}; motivational interviewing, which operationalises the relational and process behaviours that facilitate or block change\cite{miller2013}; and evidence-based accounts of the therapeutic relationship, which identify the therapist behaviours that empirically predict outcome\cite{fluckiger2018,elliott2018,norcross2019}. This approach is inspired by the MITI behavioural counts\cite{moyers2014}, but focuses on observable behaviour rather than on intention or on the exact wording used. The catalogue is organised in three families, each corresponding to a distinct mechanism by which an Advisor may undermine the support process.

\paragraph{Reactance-producing expressions.} This family captures Advisor behaviours that threaten the Advisee's perceived autonomy, thereby eliciting resistance rather than engagement\cite{rains2013,dillard2005,brehm1966}. In motivational interviewing, avoiding language that elicits reactance is treated as a foundational relational skill and is central to how sustain talk is generated or reduced in the encounter\cite{miller2013}. In acute crisis specifically, generic reassurance and formulaic empathy are widely regarded as counterproductive, because they leave the person feeling unheard rather than supported and short-circuit the specific validation that stabilisation requires\cite{wang2024,linehan1997}.

\begin{itemize}
\item \emph{False understanding} --- the Advisor asserts understanding of the Advisee's feelings in a generic, unjustified way instead of demonstrating understanding through specific reflection of what has just been said. This departs from empathy as an evidence-based relational element, where the empathic act is not the declaration but the accurate reflection\cite{elliott2018,rogers1957}.
\item \emph{Premature trust} --- the Advisor solicits, asserts or demands trust rather than allowing it to be earned through consistent behaviour over successive turns. This inverts the direction of bond formation in the working alliance, which is empirically built turn by turn rather than declared\cite{bordin1979,fluckiger2018}.
\item \emph{False reassurance} --- the Advisor offers premature or generic reassurance that dismisses the Advisee's concerns. Evidence-based crisis support prioritises acknowledging distress and promoting calming through specific validation rather than through generic reassurance\cite{wang2024,hobfoll2007,linehan1997}.
\end{itemize}

\paragraph{Process failures.} This family captures violations of the temporal sequence that evidence-based crisis support requires. Psychological first aid and crisis intervention converge on a listen-before-solving sequence in which stabilisation, rapport and exploration precede information-giving and action planning\cite{roberts2005,wang2024,hobfoll2007}. Moving prematurely to solutions is a specific instance of the righting reflex that motivational interviewing identifies as a central obstacle to the change process\cite{miller2013}, and it also runs against the client-centred principle that constructive change is preceded, not caused, by the therapist's attempt to solve the problem\cite{rogers1957}.

\begin{itemize}
\item \emph{Premature problem-solving} --- the Advisor jumps to solutions, advice or action plans before the Advisee's situation has been explored. This is the righting reflex as operationalised in motivational interviewing and is empirically associated with poorer engagement\cite{miller2013}.
\item \emph{Rushing past emotion} --- the Advisee's preceding turn contained an emotionally loaded disclosure and the Advisor moves past it with a new question, solution or topic shift without first acknowledging the stated emotion or giving it space. This closes down the emotion-processing that emotion-focused approaches and DBT identify as essential to constructive change, and inverts the crisis-support sequence in which validation precedes redirection\cite{linehan1997,greenberg2015}.
\end{itemize}

\paragraph{Boundary and scope.} This family captures behaviours that collapse the frame of the supportive interaction. A supportive encounter operates within an implicit scope: the helper attends to the person's concerns, respects the arc of the conversation and does not take over tasks that belong to the person. In trauma and crisis contexts, holding this frame is itself protective, because loss of boundaries can reproduce the loss of control the person is already experiencing\cite{herman1992}. In structured therapeutic approaches, closing the encounter is also a defined task with a specific function, not an incidental ending\cite{beck2011}.

\begin{itemize}
\item \emph{Fails to close} --- the Advisor cannot let the conversation move toward an ending. It adds another probing question or pushes to continue even when the Advisee signals that they are winding down or finished\cite{beck2011}.
\item \emph{Scope overreach} --- the Advisor oversteps its supporting role and tries to go above and beyond by taking work out of the hands of the Advisee. This endangers the space of work of the person seeking help; in crisis and trauma contexts, preserving that space is itself protective\cite{herman1992}.
\end{itemize}

We also monitor turn length. Acute stress narrows attention and impairs prefrontal executive function, reducing the working-memory resources available to process new input\cite{arnsten2009,shields2016}, so Advisor verbosity is a specific liability in this context. The behavioural counter identifies as long any Advisor turns exceeding a fixed 160-word threshold; note that 89\% of Advisor turns in this corpus exceed it, so this count is best read as an indicator of near-uniform verbosity rather than as a discriminating measure. We also compute the Advisor-to-Advisee word ratio; values greater than one indicate that the Advisor talks more than it listens over the course of the encounter.

\subsection{Scenario and task specification}

Scenarios are an important part of the simulation, both to take full advantage of the multi-turn interactions and to control for the level of adjustment of the support provided depending on different psychological profiles, as in real situations. The scenario should be designed jointly with the persona specification and should inform which Auditors are relevant. For this paper we chose the scenario where the Advisee seeking help has received the distressing news that a close relative has been diagnosed with dementia that same day, in a context of no social or family support. The scenario and task specification are given in Supplementary Table 1.

\subsection{Statistics and reproducibility}

All analyses are based on seven Advisees per Advisor, without within-cell replication, and are therefore exploratory; effect sizes with confidence intervals are the primary quantity and $p$ values are secondary. Because $n = 7$, exact and distribution-free methods are used throughout rather than large-sample approximations. Confidence intervals for standardised mean differences are exact, obtained by inverting the noncentral $t$ distribution. Confidence intervals for correlations in Study 1 come from a cluster bootstrap ($B = 10{,}000$) resampling Advisees with replacement, since the 70 construct instances per instantiating model are nested within seven Advisees and are not independent. Confidence intervals for means on bounded scales (TENS-Life 1--7, CEISS 0--36, MITI globals 1--5), for behaviour counts and for the composite score use a percentile bootstrap ($B = 10{,}000$), because a $t$-based interval would extend beyond the scale limits or below zero.

Normality of the seven paired differences behind each pre--post contrast was assessed with Shapiro--Wilk; normality was not rejected in any of the twelve contrasts (all $p \geq .10$), but because this test has low power at $n = 7$ every contrast is also reported with the distribution-free Wilcoxon signed-rank test. All between-Advisor comparisons are one-way repeated-measures analyses blocking on Advisee, since the same seven Advisees appear in every arm, with Friedman's rank test as a distribution-free check. Familywise error was controlled with the Holm procedure, applied within each pre--post instrument (four Advisors per family) and within each of three between-Advisor families: TENS subscales, support-quality metrics and behaviour counts. All tests are two-tailed with $\alpha = .05$.

Inter-rater agreement among Auditors is reported as two-way random-effects intraclass correlations with absolute agreement, so that systematic leniency differences between Auditors count against reliability rather than being partialled out.

\subsection{Use of large language models in manuscript preparation}

Large language models were used during the preparation of this manuscript for language editing and for assistance in writing analysis code. No large language model was used to generate scientific content, to interpret results or to draft the arguments presented. All authors take full responsibility for the content of the manuscript.

\subsection{Ethics}

This study involved no human or animal participants, no human material and no human data. All Advisees were synthetic and all conversations were generated between language models. On this basis the authors determined that ethical approval was not required, and no institutional review was sought.

\section*{Data availability}

The conversation transcripts, Advisee profile specifications and per-conversation Auditor scores generated and analysed in this study are available at \url{https://doi.org/10.82186/9aajj-vq030}. No human-participant data were collected.

\section*{Code availability}

The code used to generate the conversations, instantiate the Advisee profiles, run the Auditors and compute the behavioural counter, together with the statistical analysis scripts, is available at \url{https://doi.org/10.82186/9aajj-vq030}. Item text for the NEO-FFI and the CHIP inventory has been removed from the released Advisee prompt templates and replaced with placeholders, since these instruments are commercially licensed and their items cannot be redistributed; they are available from PAR and Multi-Health Systems respectively. All other instrument content is included.

\section*{Acknowledgments}

R.A.C. and P.A.F. are funded by the Leverhulme Centre for the Future of Intelligence, Leverhulme Trust. The funder played no role in study design, data collection, analysis and interpretation of data, or the writing of this manuscript.

\section*{Author contributions}

R.A.C. and R.R.-C. conceived the study. P.A.F., R.R.-C. and R.A.C. designed the four studies. P.A.F. developed the conversational platform on which the synthetic Advisees, Advisors and Auditors interact, implemented the AAA framework and the behavioural counter, generated and curated the full conversation corpus, and designed and conducted the statistical analyses. R.R.-C. specified the Advisee personality profiles, selected and adapted the Auditor instruments, defined the behavioural counter categories from the clinical literature and led the psychological interpretation of the results. R.A.C. developed the conceptual framing of the roles, contributed the self-determination measures, secured the resources for the work and supervised the project. All authors revised the manuscript critically and approved the final version.

\section*{Competing interests}

P.A.F., R.R.-C. and R.A.C. declare no financial or non-financial competing interests. No author has a financial or advisory relationship with any of the model providers evaluated.

\bibliographystyle{unsrtnat}
\bibliography{refs}

\end{document}